\documentclass[psfig]{aa}
\usepackage{graphics}
\usepackage{psfig}

\begin{document}

\title {Multiwavelength optical observations of chromospherically
active binary systems} \subtitle {V. FF~UMa (2RE~J0933+624):
 a system with orbital period variation
\thanks{Based on observations collected with the 2.2~m telescope
at the Centro Astron\'omico Hispano Alem\'an (CAHA) at Calar
 Alto (Almer\'{\i}a, Spain), operated jointly
by the Max-Planck Institut f\"{u}r Astronomie and the Instituto de
Astrof\'{\i}sica de Andaluc\'{\i}a (CSIC);
with the Nordic Optical Telescope (NOT),
 operated on the island of La Palma jointly by Denmark, Finland,
 Iceland, Norway and Sweden, in the Spanish Observatorio del
 Roque de Los Muchachos of the Instituto de Astrof\'{\i}sica de Canarias;
 with the 2.1~m Otto Struve Telescope at McDonald Observatory of
 the University of Texas at Austin (USA) and with Hobby-Eberly Telescope,
which is a joint project of the University of Texas at Austin, the
Pennsylvania State University, Stanford University,
Ludwig-Maximilians-Universit\"{a}t M\"{u}nchen, and
Georg-August-Universit\"{a}t G\"{o}ttingen. },\thanks{Tables 8 and 9 are
only available in electronic form via http://www.edpsciences.org} }

\titlerunning{FF~UMa (2RE~J0933+624: orbital period variation)}

\author{
M.C.~G\'alvez\inst{1,2}
\and D.~Montes\inst{1}
\and M.J.~Fern\'{a}ndez-Figueroa\inst{1}
\and E.~De Castro\inst{1} \and M.~Cornide\inst{1}
}

\offprints{M.C.~G\'alvez}
\mail{mcz@astrax.fis.ucm.es}

\institute{
Departamento de Astrof\'{\i}sica,
Facultad de Ciencias F\'{\i}sicas,
 Universidad Complutense de Madrid, E-28040 Madrid, Spain\\
E-mail: mcz@astrax.fis.ucm.es
\and Department of Astronomy, University of Florida, Gainesville, FL 32611, USA
}

\date{Received .. .. 2006 / Accepted .. .. 2007}

\abstract
{This is the fifth paper in a series aimed at studying the chromospheres of
 active binary systems using several optical spectroscopic indicators
 to obtain or improve orbital solution and fundamental stellar parameters.}
{We present here the study of \object{FF~UMa (2RE~J0933+624)}, a
  recently discovered, X-ray/EUV selected, active
binary with strong H$\alpha$ emission. The objectives of this work are,
 to find orbital solutions and define stellar parameters from precise radial velocities and carry out an extensive study of
 the optical indicators of chromospheric activity.}
{We obtained high resolution echelle spectroscopic
 observations during five observing runs from 1998 to 2004.
 We found radial velocities by cross
 correlation with radial velocity standard stars to achieve
 the best orbital solution. We also measured rotational velocity
 by cross-correlation techniques and have studied the kinematic by galactic
 space-velocity components ($U$, $V$, $W$) and Eggen criteria.
 Finally, we have determined the chromospheric contribution in
 optical spectroscopic indicators, from Ca~{\sc ii} H $\&$ K to
 Ca~{\sc ii} IRT lines, using the spectral subtraction technique.}
{We have found that this system presents an orbital period variation,
 higher than previously detected in other RS~CVn systems. We determined
 an improved orbital solution, finding a circular orbit with a
 period of 3.274 days. We derived the stellar parameters, confirming
 the subgiant nature of the primary component ($M_{\rm P}$ = 1.67
 $M_{\odot}$ and $R\sin{i}_{\rm P}=2.17$ $R_{\odot}$) and
 obtained rotational velocities ($v\sin{i}$), of
 33.57$\pm$0.45 km s$^{-1}$ and
 32.38$\pm$0.75 km~s$^{-1}$ for the primary and secondary
 components respectively.
 From our kinematic study, we can deduce its membership to
  the Castor moving group.
 Finally, the activity study has given us a
 better understanding of the possible mechanisms that
 produce the orbital period variation.}
{}


\maketitle

\begin{table*}
\caption[]{Observing log
\label{tab:obslog}}
\begin{flushleft}
\scriptsize
\begin{center}
\begin{tabular}{ccccccccccccccc}
\noalign{\smallskip}
\hline \hline
\noalign{\smallskip}
\multicolumn{3}{c}{2.1m-Sandiford 1998/01} &
\multicolumn{3}{c}{9.2m McDonald 2000/01} &
\multicolumn{3}{c}{2.2m-FOCES 2002/04} &
\multicolumn{3}{c}{2.2m-FOCES 2004/04} &
\multicolumn{3}{c}{NOT-SOFIN 2004/04}  \\
\hline

\noalign{\smallskip} \scriptsize Day  & \scriptsize UT &  \scriptsize $S/N$
& \scriptsize Day  & \scriptsize UT &  \scriptsize $S/N$ & \scriptsize Day
& \scriptsize UT &  \scriptsize $S/N$ & \scriptsize Day  & \scriptsize UT &
\scriptsize $S/N$ & \scriptsize Day  & \scriptsize UT &  \scriptsize $S/N$
\\
 &  & \scriptsize (H$\alpha$) &
 &  & \scriptsize (H$\alpha$) &
 &  & \scriptsize (H$\alpha$) &
 &  & \scriptsize (H$\alpha$) &
 &  & \scriptsize (H$\alpha$)
\scriptsize
\\
\noalign{\smallskip}
\hline
\noalign{\smallskip}
\noalign{\smallskip}
13 & 10:27 &  146  & 19 & 06:12 &  233  & 22 & 19:54 &  118 &  31 & 23:36 & 71
& 2 & 21:53 & 225 \\
14 & 9:58 &  203   & 23 & 06:34 &  263  & 23 & 19:41 &  200 &  2  & 19:38 & 63
& 2 & 22:02 & 180 \\
15 & 10:07 &  98   & 24 & 07:29 &  83  & 24 & 19:45 &  140  &  3  & 19:23 & 87
& 4 & 00:56 & 243 \\
16 & 10:19 &  139   & 25 & 06:24 &  146  & 25 & 21:55 &  146 &  3 & 23:22 & 74
& 5 & 21:33 & 227 \\
17 & 10:23 &  95   & 26 & 10:12 &  236  &   &  &  &   4 & 02:46 & 59 & & & \\
18 & 11:04 &  134   &  &  & & & &  &    4 & 19:18 & 95 &  & &\\
19 & 11:00 &  146   &  &  & & & &  &    4 & 22:23 & 117 &  & &\\
20 & 10:27 &  83   &  &  & & & &   &    5 & 01:43 & 80 &  & &\\
21 & 10:52 &  77   &  &  & & & &   &    5 & 19:23 & 56 & & &\\
22 & 12:57 &  115  &  &  & & & &  &     5 & 22:29 & 122 &  & &\\
   &       &       &  &  & & & &  &     6 & 02:14 & 130 &  & &\\
   &       &       &  &  & & & &  &     6 & 19:20 & 117 &  & &\\
   &       &       &  &  & & & &  &     7  & 01:32 & 93 &  & &\\
\noalign{\smallskip}
\noalign{\smallskip}
\hline
\end{tabular}
\end{center}
\end{flushleft}
\end{table*}

\section{Introduction}

This paper is a continuation of our ongoing project aimed at studying
the chromospheres of active binary systems using multiwavelength
optical observations. These observations provide
the information for several optical spectroscopic
features that are formed at different heights in the chromosphere (see
Montes et al. 1997, Paper~I; Montes et al. 1998, Paper~II; Montes et al.
2000, Paper~III; G\'alvez et al. 2002, Paper~IV).
In addition to studying stellar activity, our high resolution spectroscopic
observations allow us to determine radial
velocities and obtain and improve fundamental stellar parameters.
While several systems have been studied, this is the first time
we have found an orbital period variation, giving us new clues into
the understanding of activity-orbit relation.
  When combined with other examples,
the study of this type of system could help us
 understand how orbital dynamics are affected by physical
 processes intrinsic to the binary system (Lanza 2006).

We focus here on the X-ray/EUV selected chromospherically active binary
FF~UMa (2RE~J0933+624, HD~82286, SAO~14919). It is an SB2 system with  $V
= 8.35$ mag.  First classified by Jeffries et al. (1995) as two G5V or
G5V/G5IV stars, it was reclassified by Henry et al. (1995) and
 Strassmeier et al. (2000) as an K0IV/K0IV.

Henry et al. (1995) reported a photometric period of 3.270 days
 obtained from a periodogram analysis derived
 from photometric data, and  estimated a
rotational velocity $v\sin{i}$ = 35 km s$^{-1}$ for both components.
Jeffries et al. (1995) obtained an orbital period of 3.28 days from
 15 radial velocity measurements and suggested an eccentricity less than
 0.18. Their estimated value of $R\sin{i}$ was in agreement with  a
 subgiant primary.
Strassmeier et al. (2000) found a rotational period of 3.207 days and
 photometric variation amplitudes of $\Delta$$V$ $\approx$ 0.15 mag.
All previous authors have reported that this star presents very
 strong chromospheric activity and the H$\alpha$
emission line is detected above the continuum for both components.

In this paper, we present high-resolution echelle spectra of this system.
We measured radial velocities using the cross-correlation
technique and obtained an orbital period variation during
  11 years of observations.  In spite of this variation, we achieved a
 good orbital solution, finding that the mean orbital period is similar
 to the photometric one, indicating synchronous rotation.

In addition, we applied the spectral
subtraction technique to study the chromospheric excess emission
in the Ca~{\sc ii} H \& K, Ca~{\sc ii} IRT, H$\alpha$ and
other Balmer lines of the primary and secondary components of the system.
Preliminary results for this system can be found in G\'alvez (2005); and
G\'alvez et al. (2006, 2007).

In Sect.~2 we give the details of our observations and data reduction.
In Sect.~3 we discuss the nature of the orbital period variation
 and give the orbital solution of the binary system.
 In Sect.~4 we give the derived stellar and
 kinematic parameters. The behavior of the different chromospheric
activity indicators is described in Sect.~5.
Finally, in Sect.~6 we present our conclusions.

\section{Observations and data reduction}

We obtained high resolution echelle spectra of FF UMa during five observing
runs from 1998 to 2004:

{1)} {2.1~m-SANDIFORD, McDonald Obs., 1998/01} \\
During this observing run, which extended from 12 to 21 January 1998,
 we used the 2.1~m Otto
Struve Telescope at McDonald Observatory Texas (USA) with the
Sandiford Cassegrain Echelle Spectrometer (SCES), equipped with a
  1200x400 pixel CCD detector. The
wavelength range covers from 6400 to 8800 \AA \ in 31 orders. The
reciprocal dispersion ranges from 0.06 to 0.08 \AA/pixel and the
spectral resolution, determined as the full width at half maximum (FWHM)
 of the arc comparison lines, from 0.13 to 0.20 \AA. In the fourth night,
 the central wavelength was changed to include the Na~{\sc i} D$_{1}$,
D$_{2}$ (5889.95, 5895.92 \AA) and He~{\sc i} D$_{3}$ (5876 \AA) lines.
 Therefore the wavelength range changed to 5600-7000 \AA.

{2)} {9.2~m-HET, McDonald Obs., 2000/01}\\
 We used the 9.2~m Hobby-Eberly Telescope (HET)
 and the medium resolution spectrograph UFOE (Upgraded Fiber Optic Echelle)
 equipped with a 1200x400 pixel CCD detector, located at McDonald
 Observatory Texas (USA) on 22 to 24 January 2000.
 The wavelength range covers from 4400 to 9150 \AA \ in 26 orders.
 The reciprocal
 dispersion ranges from 0.06 to 0.17 \AA/pixel and the spectral
 resolution (FWHM) ranges from 0.14 to 0.42 \AA.

{3) and 4)} {2.2~m-FOCES, CAHA, 2002/04 and 2004/04} \\
 We utilized the Fibre Optics Cassegrain Echelle
Spectrograph (FOCES) (Pfeiffer et al. 1998) with
a 2048x2048 24$\mu$ SITE$\#$1d CCD detector on
 the 2.2~m telescope at the German Spanish Astronomical Observatory (CAHA)
(Almer\'{\i}a, Spain) to obtain spectra between
 22 to 26 April 2002 and from 29 March to 7 April 2004.
The wavelength range
covers from 3450 to 10700 \AA\ in 112 orders. The reciprocal
dispersion ranges from 0.04 to 0.13 \AA/pixel and the spectral
resolution (FWHM) ranges from 0.08 to 0.35 \AA.

{5)} {2.56~m-NOT-SOFIN, Roque de los Muchachos Obs., 2004/04}\\
We used the 2.56~m Nordic Optical Telescope (NOT) located at the
 Observatorio del Roque de los Muchachos (La Palma, Spain) on
 2 to 6 April 2004. We used The Soviet Finnish High Resolution
 Echelle Spectrograph (SOFIN) with an echelle grating
 (79 grooves/mm), ASTROMED-3200 camera and a 2052x2052 pixel 2K3EB PISKUNOV1
 CCD detector. The wavelength range covered from 3545 to 10120 \AA \
 in 42 orders. The reciprocal dispersion ranges from 0.033 to 0.11 \AA/pixel
 and the spectral resolution (FWHM) from 0.14 to 0.32 \AA.
 We note that we had some problems with the wavelength calibration,
 during the arc lamp spectra exposures, and as a consequence,
   could not rely on the absolute wavelength calibration
 of the spectra taken in this observing run.
Therefore, these spectra have not been used to determine radial
velocities, although we have used them for the remaining analysis.

In Table~\ref{tab:obslog} we present the observing log.
For each observation we list date, UT,
and the signal to noise ratio ($S/N$) obtained in the H$\alpha$ line region.

We extracted spectra using the standard
reduction procedures in the
IRAF\footnote{IRAF is distributed by the National Optical Observatory,
which is operated by the Association of Universities for Research in
Astronomy, Inc., under contract with the National Science Foundation.}
 package (bias subtraction,
flat-field division and optimal extraction of the spectra).
The wavelength calibration was obtained by taking
spectra of a Th-Ar lamp.
Finally, we normalized the spectra by
a polynomial fit to the observed continuum.

\section{Orbital period variation}

\begin{table}
\caption[]{Radial velocities
\label{tab:vr}}
\scriptsize
\begin{tabular}{llrrrrrrrrrrrllrrrrrrrr}
\noalign{\smallskip}
\hline \hline
\noalign{\smallskip}
 Obs. & HJD & \tiny $S/N$ & \multicolumn{1}{c}{Primary} & \multicolumn{1}{c}{Secondary} \\
\noalign{\smallskip}
    & 2400000+& (H$\alpha$)& $V_{\rm hel}$ $\pm$ $\sigma_{V}$ &
            {\rm $V_{\rm hel}$} $\pm$ $\sigma_{V}$   \\
\noalign{\smallskip}
    &  & & \tiny (km s$^{-1}$) & \tiny (km s$^{-1}$)\\
\noalign{\smallskip}
\hline
\noalign{\smallskip}
 Jef(95)$^{1}$ & 49054.591 &- & 23.5 $\pm$ 4.0 & -35.9 $\pm$ 9.0 \\
 Jef(95)$^{1}$ & 49054.634 &- & 26.8 $\pm$ 3.0 & -30.0 $\pm$ 9.0 \\
 Jef(95)$^{1}$ & 49054.675 &- & 22.6 $\pm$ 3.0 & -45.0 $\pm$ 5.0 \\
 Jef(95)$^{1}$ & 49055.378 &- & 20.0 $\pm$ 3.0 & -53.3 $\pm$ 3.0 \\
 Jef(95)$^{1}$ & 49055.423 &- & 19.5 $\pm$ 3.0 & -47.6 $\pm$ 4.0 \\
 Jef(95)$^{1}$ & 49056.425 &- & -25.1 $\pm$ 5.0 & 48.7 $\pm$ 3.0 \\
 Jef(95)$^{1}$ & 49056.486 &- & -20.6 $\pm$ 4.0 & 58.8 $\pm$ 3.0 \\
 Jef(95)$^{1}$ & 49056.543 &- & -28.1 $\pm$ 4.0 & 57.8 $\pm$ 3.0 \\
 Jef(95)$^{1}$ & 49056.600 &- & -29.4 $\pm$ 4.0 & 57.3 $\pm$ 3.0 \\
 MCD98 & 50826.935  & 146 &25.80 $\pm$ 1.92 & -62.05 $\pm$ 4.75 \\
 MCD98 & 50827.915  & 203 &-11.09 $\pm$ 6.26 &  26.23  $\pm$ 6.62 \\
 MCD98 & 50828.922  & 98 & -22.16 $\pm$ 1.71 & 49.39 $\pm$ 4.58 \\
 MCD98 & 50829.929  & 139 & 25.95  $\pm$ 1.86 & -60.06 $\pm$ 4.29 \\
 MCD98 & 50830.932  & 95 &-1.88 $\pm$ 6.25 &   - \\
 MCD98 & 50831.961  & 134 & -28.27 $\pm$ 1.72 & 54.33 $\pm$ 4.59 \\
 MCD98 & 50832.958  & 146 & 17.88 $\pm$ 1.63 & -50.03 $\pm$ 6.42 \\
 MCD98 & 50833.935  & 83 & 9.34 $\pm$ 4.51 & -33.52  $\pm$ 4.72 \\
 MCD98 & 50834.953  & 77 &  -29.42 $\pm$ 1.78 & 55.83 $\pm$ 4.90 \\
 MCD98 & 50836.039  & 115 & -4.99  $\pm$ 6.20 & - \\
 HET00 & 51561.943  & - & -27.96 $\pm$ 2.60 & 61.98 $\pm$ 2.80 \\
 HET00 & 51562.759  & 233 &  -3.74 $\pm$ 6.51 & - \\
 HET00 & 51562.837  & - &  -2.75 $\pm$ 6.50 & -\\
 HET00 & 51566.774  & 263 &  24.78 $\pm$ 2.66 & -62.86 $\pm$ 2.81 \\
 HET00 & 51566.797  & - & 25.66 $\pm$ 2.66 & -65.02 $\pm$ 2.81\\
 HET00 & 51567.812  & 83 &  -0.97 $\pm$ 6.21 & - \\
 HET00 & 51568.767  & 146 & -31.61 $\pm$ 2.29 & 58.01 $\pm$ 2.70 \\
 HET00 & 51569.925  & 236 &  21.76 $\pm$ 2.64 & -58.45 $\pm$ 2.53 \\
 FOCES02 & 52387.329  & 118 & -31.89 $\pm$ 2.33 & 54.19 $\pm$ 4.49 \\
 FOCES02 & 52388.320  & 200 & -2.24 $\pm$ 4.13 & -  \\
 FOCES02 & 52389.323  & 140 &24.44 $\pm$ 2.56 & -54.03 $\pm$ 4.62 \\
 FOCES02 & 52390.414  & 146 &-24.46 $\pm$ 2.63 & 47.34 $\pm$ 4.93 \\
 FOCES04 & 53096.4836 & 71 & 26.98 $\pm$ 2.58 & -63.76 $\pm$ 6.27  \\
 FOCES04 & 53098.3186 & 63 &-32.32 $\pm$ 2.06 & 58.08 $\pm$ 6.46  \\
 FOCES04 & 53099.3077 & 87 & 10.94 $\pm$ 3.24 & -41.79 $\pm$  8.17  \\
 FOCES04 & 53099.4736 & 74 & 19.50 $\pm$ 2.98 & -51.18 $\pm$ 5.88  \\
 FOCES04 & 53099.6158 & 59 & 23.20 $\pm$ 2.31 & -58.85 $\pm$ 6.54  \\
 FOCES04 & 53100.3048 & 95 & 18.38 $\pm$ 4.44 & -41.31 $\pm$ 8.71  \\
 FOCES04 & 53100.4329 & 117 & 10.31 $\pm$ 4.15 & -30.65 $\pm$ 8.26  \\
 FOCES04 & 53100.5718 & 80 &  0.24 $\pm$ 4.41 & - \\
 FOCES04 & 53101.3082 & 56 &-31.11 $\pm$ 2.03 & 55.25 $\pm$ 5.04   \\
 FOCES04 & 53101.4374 & 122 &-31.81 $\pm$ 2.44 & 58.15 $\pm$ 5.54  \\
 FOCES04 & 53101.5932 & 130 &-32.02 $\pm$ 2.18 & 58.24 $\pm$ 5.44  \\
 FOCES04 & 53102.3061 & 117 & -3.95 $\pm$ 4.25 &  -  \\
 FOCES04 & 53102.5640 & 93 & 11.02 $\pm$ 3.46 &-40.39 $\pm$ 8.89   \\
\noalign{\smallskip}
\hline
\noalign{\smallskip}
\end{tabular}
\\
$^{1}$ JEF(95): Jeffries et al. (1995)
\end{table}

\subsection{Radial velocities}
We determined the heliocentric radial velocities by making use of
 cross-correlation technique (see Paper~IV). The spectra of the
 target were cross-correlated order by order, using the routine
{\sc fxcor} in IRAF, against spectra of radial velocity standards with
similar spectral type taken from Beavers et al. (1979).
We derived the radial velocity for each order from the position of peak of the
cross-correlation function (CCF) and calculated the uncertainties
 based on the fitted peak height and the antisymmetric noise as
described by Tonry \& Davis (1979).
As FF~UMa is an SB2 system we note two peaks in the CCF, associated
 with the two components, and fit each one separately. When the
 components are too close, we used deblending fits.
 It is worth mentioning that the uncertainties returned by {\sc fxcor}
 for SB2 binaries are overestimated; when
fitting each star, the presence of the other will increase the
antisymmetric noise, thereby biasing the error.

\begin{figure}
\begin{center}
{\psfig{figure=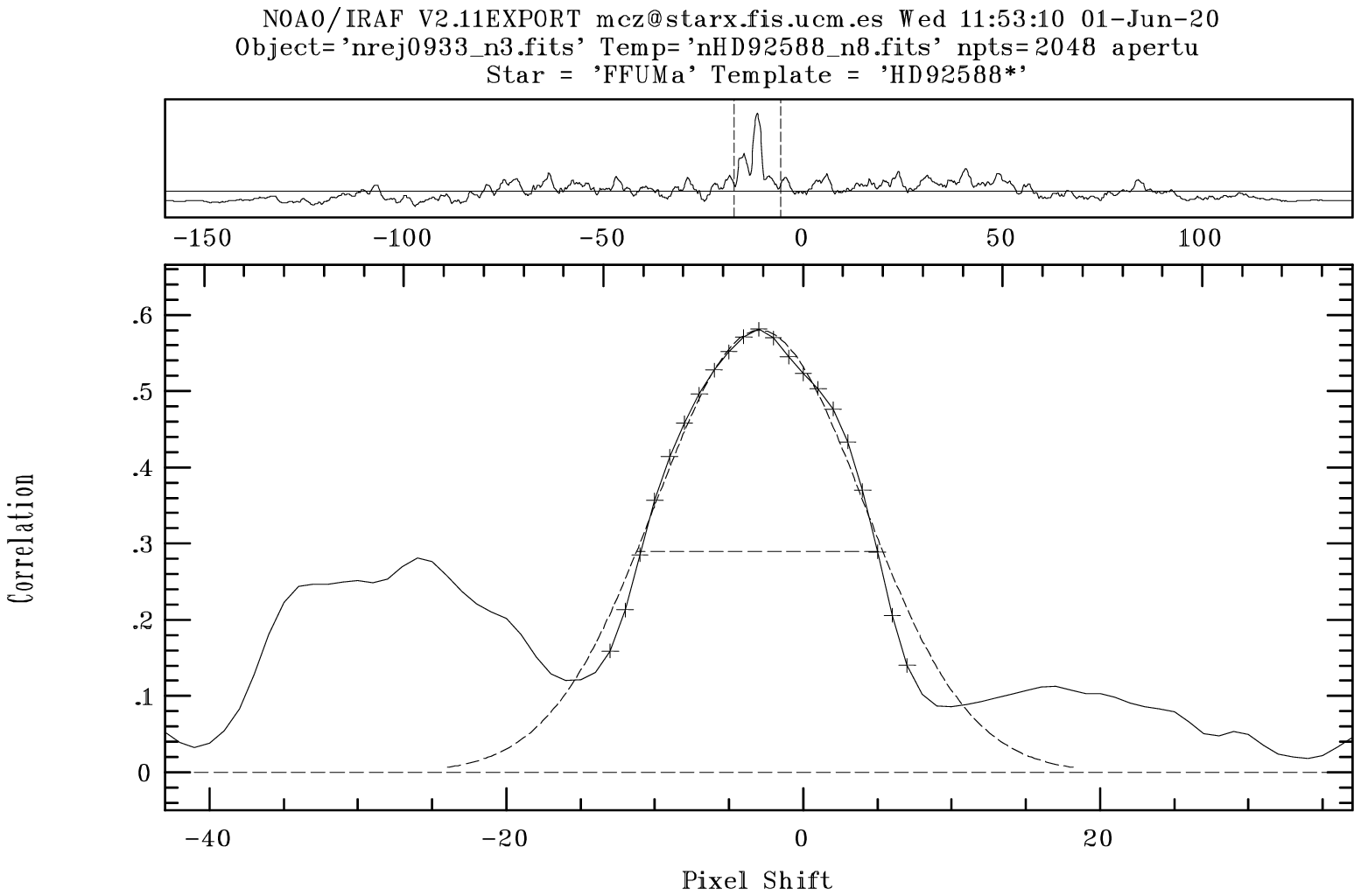,bbllx=86pt,bblly=257pt,bburx=534pt,bbury=523pt,height=6.6cm,width=8.6cm,clip=}}
\caption[]{ CCF of FF~UMa (2RE~J0933+624) in FOCES04 observing run.
 Irregular profiles can be seen in the two peaks. These
 irregularities can produce significative errors in radial velocity
 determination.
\label{fig:fx}}
\end{center}
\end{figure}

\begin{figure}
\begin{center}
{\psfig{figure=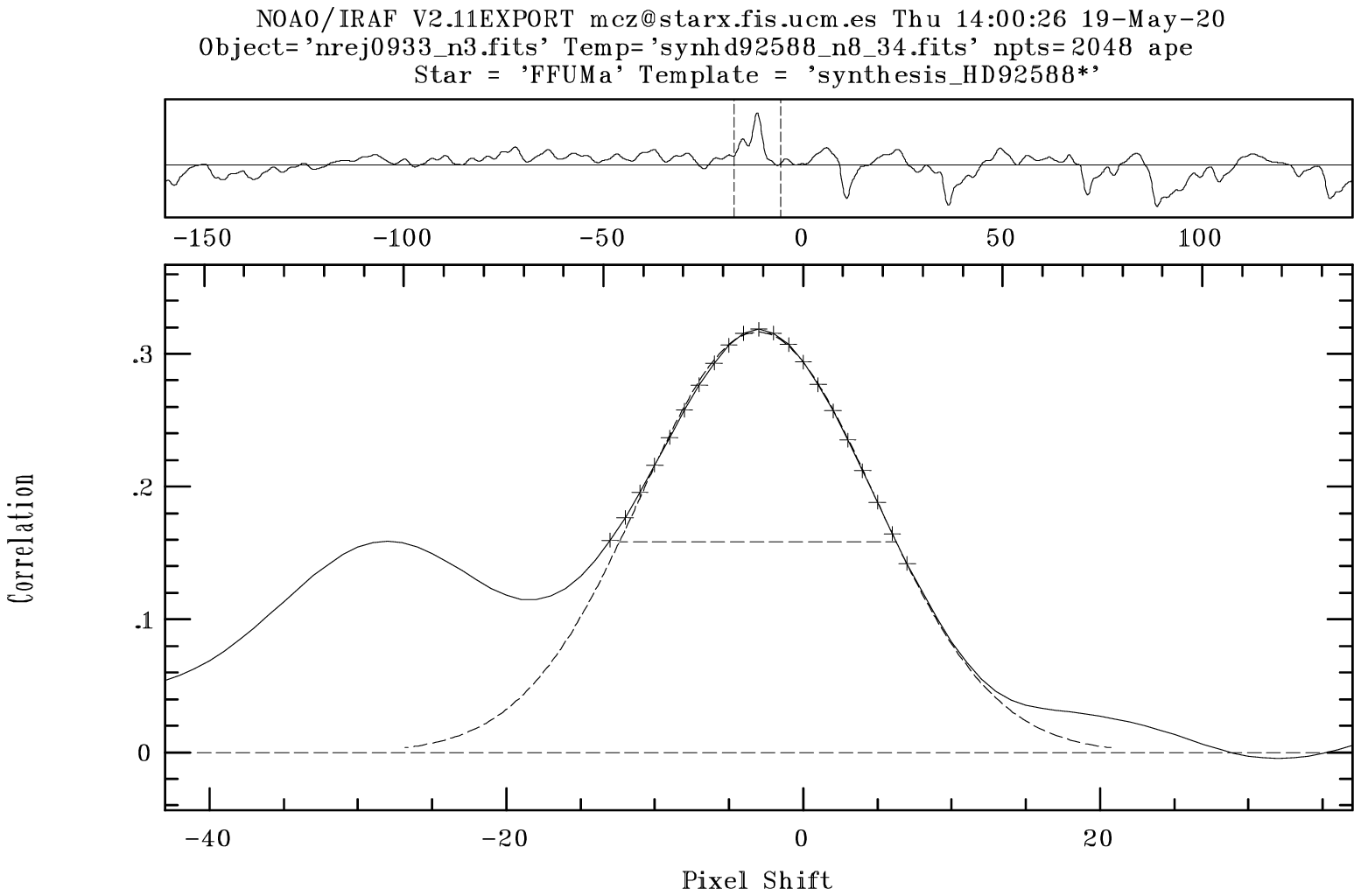,bbllx=86pt,bblly=257pt,bburx=534pt,bbury=523pt,height=6.6cm,width=8.6cm,clip=}}
\caption[] {CCF of FF~UMa (2RE~J0933+624) in FOCES04 observing run
obtained when
 we broad the standard star to FF~UMa rotational velocity.
 Irregular profiles become smoother and can be
 fitted with a Gaussian.
\label{fig:fx2}}
\end{center}
\end{figure}

 As Fig.~\ref{fig:fx} shows, the irregular profiles of
the CCF (double peaks and asymmetries) can produce significant errors in
radial velocity measures.
 These irregularities may
 come from photospheric activity features on the stellar
 surface of both components that disturb the profile of the
 photospheric lines and induce variations in the peak of the CCF.
 However, this behavior may also be due to the difference in rotational
 velocity ($v\sin{i}$)
 between the problem and radial velocity star.
 When the spectrum of the standard star
 was broadened to the same rotational velocity of FF~UMa
 ($v\sin{i}$ $\approx$ 30 km s$^{-1}$) the profiles of CCF become
 smoother and could be fit with
 a Gaussian profile, see Fig.~\ref{fig:fx2}.
Therefore all the radial velocities given in this paper have been
calculated by cross-correlation with this rotational broadened spectrum of
the standard star.

In Table~\ref{tab:vr} we list, for each spectrum, the heliocentric radial
velocities ($V_{\rm hel}$) and their associated errors ($\sigma_{V}$)
obtained as weighted means of the individual values deduced for each
 order in the spectra.
Those orders which contain chromospheric features and prominent telluric
lines have been excluded when determining the mean velocity.

\subsection{$T_{\rm conj}$ variations}

 With 35 radial velocity data from our measures and nine
 from Jeffries et al. (1995) (see Table~\ref{tab:vr}), we
 computed the orbital solution of this system.
 Although we obtained good results when
  we fit orbital solution for each observing run data separately,
 some orbital parameters changed from one epoch to another.
 When we tried to fit the orbital solution
 with all the data, we could not find any satisfactory result.

\begin{figure*}
{\psfig{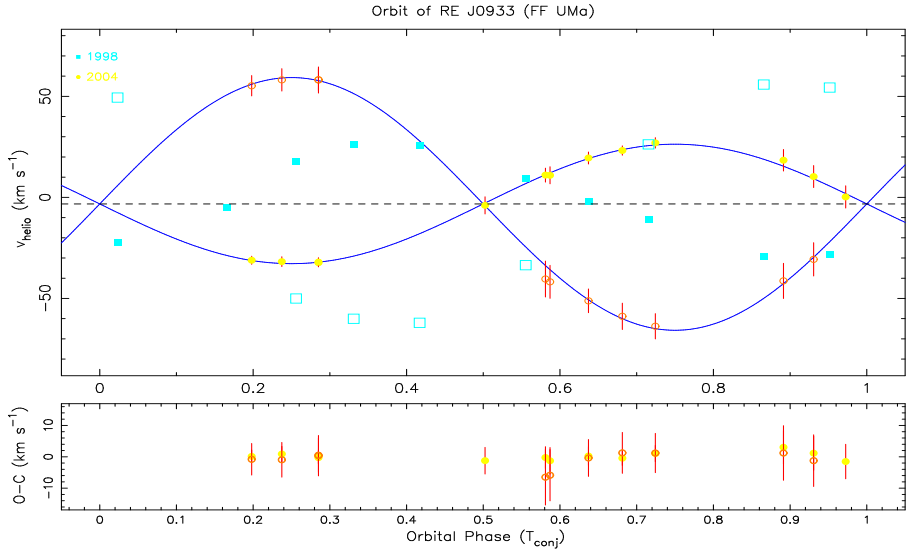}}
\caption[] {Example of the phase shift between orbital solution of FOCES04
and
 McDonald98 observing runs. Radial velocities and orbital fit (solid line)
 of FOCES04 is plotted and radial velocities of McDonald98 are superimposed.
 Filled symbols correspond to primary and open symbols to secondary.
\label{fig:desp}}
\end{figure*}

\begin{figure*}
{\psfig{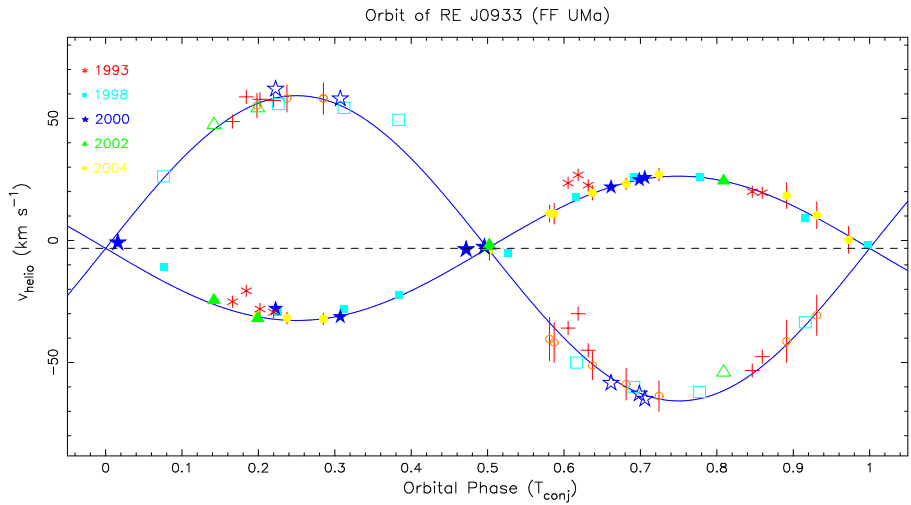}}
\caption[] {Radial velocities of all observing runs. The orbital solution
fit of
 FOCES04 observing run is plotted here (solid line) and the radial
 velocities of the rest runs (shifted in phase) are superimposed. See text
 for explanation (Sect. 3.2).
\label{fig:orb}}
\end{figure*}

 We decided to recalculate the orbital solution of each observing run,
 using the period obtained from the FOCES04 observing run
 and assuming a circular orbit (since $e$ is only $\approx$ 10$^{-2}$).
 We determined that the solutions are very similar except at time of conjunction,
  $T_{\rm conj}$, (see Fig.~\ref{fig:desp}).
 Therefore we shifted in phase the
 solutions of every run taking as standard the FOCES04 solution.
 As we can see in Fig.~\ref{fig:orb}, all data points are now
 in agreement with the orbital solution. We obtained the phase shift
 calculating the conjunction time differences between the conjunction time
 obtained in every run fit and
 the conjunction time of FOCES04 run fit,
 ($O-C$) (Observed - Calculated = $T_{\rm conj}$ difference coming from orbital
 solution fit in each run and the FOCES04 run).

\begin{table}
\caption[] {Variations of Orbital Period ($P$)
\label{tab:dpp}}
\small
\begin{center}
\begin{tabular}{lccc}
\noalign{\smallskip}
\hline \hline
\noalign{\smallskip}
Year & $T{\rm conj}$ & $O-C$ & $dP/P$ \\
\noalign{\smallskip}
    & HJD& day & \\
    & (2400000 +)& & \\
\noalign{\smallskip}
\hline
\noalign{\smallskip}
1993 & 49055.8789 & -2.0089 & 4.979x10-4\\
1998 & 50824.4023 & -1.1755 & 5.187x10-4\\
2000 & 51561.2227 & -0.9005 & 5.887x10-4\\
2002 & 52386.6758 & -0.3615 & 5.134x10-4\\
2004 & 53090.8398 & & \\
\noalign{\smallskip}
\hline
\end{tabular}
\end{center}
\end{table}

\begin{figure}
\psfig{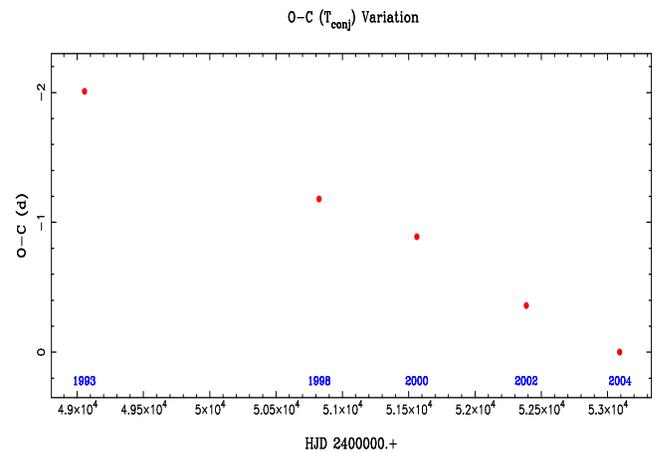}
\caption[]{The ($O-C$) (Observed - Calculated $T_{\rm conj}$) versus
  Heliocentric Julian date HJD for every observing run.
\label{fig:fig4}}
\end{figure}

In Fig.~\ref{fig:fig4}, we represent
 the temporal variation of the $O-C$ ($T_{\rm conj}$) for every
 run. The
 maximum amplitude of the variation found in our data is 2 days and
 shows a decreasing linear tendency.
We would need a longer temporal range of observations to test if the
 tendency remains linear or becomes sinusoidal as one might expect if
 there is a cycled behavior related with the activity cycle
  (see Frasca $\&$ Lanza 2005 and Sect. 3.3).

When the $O-C$ ($T_{\rm conj}$) variations are transformed to relative orbital
period variations, we find
  $dP/P$ $\approx$ 10$^{-4}$ in 11 years, that is, one order
 of magnitude higher than the largest value observed until now in HR~1099,
 (see Table~\ref{tab:dpp}).

\subsection{Discussion}
\begin{figure}
\psfig{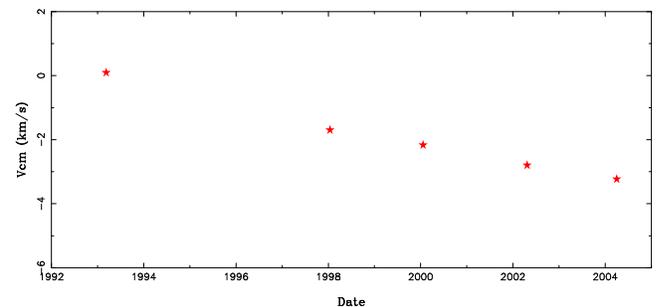}
\caption[]{Center of mass velocity, $\gamma$, obtained in each
 observing run fit versus time.
\label{fig:gamma}}
\end{figure}

To explain the behavior described above, we considered several options:

\begin{itemize}
\item First, we explore how the existence of a third distant star as a
 component of the system could modify the main orbit.
 In Fig.~\ref{fig:gamma}, we plotted the center of mass
 radial velocity, $\gamma$, obtained for each observing run, versus time.
The amplitude of variations in $\gamma$ amounted to 3 km~s$^{-1}$, over
11 years. Such differences are large enough that they are unlikely to be
due to zero-point (instrumental) differences between different runs. The
variations in $\gamma$ could indicate the presence of a third component; if
the third star is small and its period long, the reflex motion of the
binary will necessarily be small. As an example, using eq.(30) from
Cumming (2004), a third star with a mass between 0.4 and 0.6 $M_{\odot}$
and an orbital period of 20 to 40 yr in an edge-on circular orbit
 would produce a semi-amplitude $K$ on the binary between
2.1 and 3.8 km~s$^{-1}$. In addition, if the
orbit were significantly eccentric, as is often the case for
such long periods, the $K$ amplitude could be larger.
 The presence of the third component
could easily induce a change in $\gamma$ similar to that observed.
 Therefore with the present data we cannot dismiss the possibility
 that these variations are due to a third body.

\item Another explanation of our observations could be an orbital
 modulation due to the variation of activity with time,  explained
 as a consequence of cyclical variations of the quadrupole-moment of
 both components of the system during the magnetic activity cycle.
  This mechanism presented by Matese $\&$ Whirtmere (1983) and developed
 by Applegate (1992) and Lanza et al. (1998), has been used in the
 study of several RS~CVn systems such as \object{SZ~Psc}
 (Kalimeris et al. 1995), \object{RT~Lac},
 \object{RS~CVn}, \object{WW~Dra}, etc. (Lanza $\&$ Rodon\'o 1999),
 \object{XY~UMa} (Sowell et al. 2001)
 and \object{HR~1099} (Garc\'ia-\'Alvarez et al. 2003; Frasca $\&$ Lanza 2005,
 Lanza 2006).

Applegate (1992), described the initial model in which the orbital
period variation is due to the gravitational coupling of the orbit to
 changes in the quadrupole moment (rotational oblateness) of
 a magnetically active star in the system. The quadrupole moment of
 a star is determined by the rotation rate of its outer layers
 -if angular momentum is transferred to the outer layers,
 they rotate faster and the star becomes more oblate.
 The gravitational acceleration varies if the shape varies; this shape
 variation is measured by the change of the quadrupole moment of the star.
 On the contrary, if the outer layers loses angular momentum, the
 oblateness decreases.
 As the dynamo mechanism implies the qualitative shearing of magnetic
 field by differential rotation, the last should vary through the activity
 cycle. Applegate (1992) says that quantitatively, a subsurface
 magnetic field of several kilogauss
 can exert a large enough torque to transfer the angular momentum needed
 to make the observed period changes.

 Lanza et al. (1998) studied several possibilities of the process
 to explain the period variations in RS~CVn systems with different kind of
 dynamos. They showed that variations of about 100 Gauss in a poloidal magnetic
 field could produce the observed variations, while Applegate suggested
 variations in order of kilogauss. Lanza $\&$ Rodon\'o (1999) compiled 46 binary
 systems of different types (RS~CVn, WW~UMa, etc.) to evaluate the effects
 of the quadrupole moments change.

 Lanza (2005), analyzes the Applegate model predictions and the observed
 results in RS~CVn stars. He suggested that the model should be rejected
 because it fails to explain
 the orbital period variations of classical  RS~CVn close binary systems.
  The required variation of the internal differential
 rotation is too large to both agree with the observations and oppose
  turbulent dissipation. He concludes that any similar hypothesis to
 explain this phenomenon should include the effect of the Lorentz force on the
 gravitational quadrupole moment, or,
 that an entirely new theoretical framework is needed to interpret the observed
 orbital period variations in magnetically RS~CVn binaries.

  Based on the Lanza (2005) review of the Applegate model,
 Frasca $\&$ Lanza (2005) and Lanza (2006), continued
 with the characterization of the orbital period variation of HR~1099.
 They suggested that there is an interaction between the magnetic fields of
 the K1 IV subgiant (primary component),
 and the magnetic fields of the G5 V component (secondary). In the primary,
  the hydromagnetic dynamo action is maintained in the deep fast-rotating
 convective envelope, while in the secondary, the magnetic field
 comes from an outer convection
 zone with a smaller radial extension, implying that its dynamo is
 less efficient (reflected by its lower level of activity).
 They argued that the Applegate classic model could not explain
 the variation found while their assumptions of the relation between
 orbital period cycle and the activity cycle could explain the
 large variation measures. They mention, however, the need to
 verify their claim with a larger study including other systems.

 Summing up, the results found until now indicate that
 variations in the orbital period
 based on the Applegate model should be revised and that
 the strong temporal activity observable changes could reflect the
 relation between the orbital period variations and the changes
 in magnetic field and gravitational quadrupolar moment.
 The variations found in previous RS~CVn systems are about
 $dP/P$ $\approx$ 10$^{-6}-10^{-5}$ in a 7 to 109 years range.

The high level of chromospheric activity of both components of
FF~UMa (spectral types K1~IV and K0~V, see Sects. 4.2 and 5) could
imply a strong interaction between larger and more efficient
dynamos. This could explain the order of magnitude difference
between the orbital period variation of this system and the one
detected in other RS~CVn systems like HR~1099.

\item
 Finally, although the above explained activity-related period
variation is our preferred interpretation, we should mention that in the
case of the presence of a third component, there could well be changes in
the elements of the inner orbit. If the eccentricity were small but
non-zero there could be apsidal motion in the binary due to the third body
and the change in the longitude of periastron would be
seen as a change in the time of conjunction.

\end{itemize}

\subsection{Orbital solution}

As a consequence of the results in Sects. 3.2 and 3.3,
 we have computed the orbital solution of this system using radial
velocity data from the FOCES04 observing run. We chose this run
 because it has a large number of data points (13) and superior spectral
 resolution.
The radial velocity data are plotted in Fig.~\ref{fig:orb}. Solid
symbols represent the primary and open symbols represent the
secondary. Each observing run is represented with a different symbol.
 The curve represent a minimum $\chi^{2}$ fit orbit
solution. The orbit fitting code uses the {\it Numerical Recipes}
 (Press et al. 1986)
implementation of the Levenberg-Marquardt method of fitting a
non-linear function to the data, which weights each datum according
to its associated uncertainty. The program
simultaneously solves for the orbital period, $P_{\rm orb}$, the
epoch of periastron passage, $T_{\rm conj}$, the longitude of periastron,
$\omega$, the eccentricity, $e$, the primary star's radial
velocity amplitude, $K_{\rm P}$, the heliocentric center of mass
velocity, $\gamma$, and the mass ratio, $q$.  The orbital
solution and relevant derived quantities are given in
Table~\ref{tab:orb}. In this table, we
give $T_{\rm conj}$ as the heliocentric Julian date of conjunction
with the hotter star behind the cooler star, in order to adopt the
 same criteria used in previous papers.
We used this criterion
to calculate the orbital phases of all the observations reported
in this paper.

\begin{table}
\caption[]{Orbital solution
\label{tab:orb}}
\small
\begin{center}
\begin{tabular}{lccc}
\noalign{\smallskip}
\hline \hline
\noalign{\smallskip}
Element & Value & Uncertainty & Units \\
\noalign{\smallskip}
\hline
\noalign{\smallskip}
 $P_{\rm orb}$ & 3.274     & 0.054  & \small days  \\
 $T_{\rm conj}$ & 53090.84   & 0.18  & \small HJD (2400000 +) \\
 $\omega$ &  0.00   &  0.00 & degrees \\
 $e$        &     0.00 & 0.00  &  (adopted) \\
 $K_{\rm P}$  &    29.55 & 0.95  & \small km~s$^{-1}$\\
 $K_{\rm S}$  &    62.52 & 3.60  & \small km~s$^{-1}$  \\
 $\gamma$ &    -3.23 & 0.77  & \small km~s$^{-1}$  \\
 $q=M_{\rm P}/M_{\rm S}$    &    2.12  & 0.10  &   \\
\\
 $a_{\rm P}$~sin$i$  &     1.330 & 0.048  & \small 10$^{6}$~km \\
 $a_{\rm S}~$sin$i$  &     2.81 & 0.17  & \small 10$^{6}$~km \\
 $a$~sin$i$       &     4.14 & 0.18  & \small 10$^{6}$~km \\
 "            &     0.028 &         & \small AU  \\
 "            &     5.95 &         & \small R$_{\odot}$ \\
\\
 $M_{\rm P}$~sin$^{3}i$  &    0.180  & 0.024  & \small $M_{\odot}$\\
 $M_{\rm S}$~sin$^{3}i$  &     0.085 & 0.012  & \small $M_{\odot}$ \\
 f($M$)     &   0.00875 & 0.00099 & \small $M_{\odot}$ \\

\noalign{\smallskip}
\hline
\end{tabular}
\end{center}
\end{table}

This binary results in a circular orbit (adopted) with an orbital period
of about 3.274 days. Since $P_{\rm phot}$ $\approx$ 3.270 days, we can say
 that it is a synchronous system. The mass ratio of 2.12 calculated
 let us conclude that the components have a different spectral type.
The obtained parameters are in
agreement with the values reported by Jeffries et al. (1995).

\begin{table*}
\caption[]{Stellar parameters of \object{FF~UMa} \label{tab:par}}
\begin{flushleft}
\scriptsize
\begin{tabular}{l c c c  c c c c c c c }
\noalign{\smallskip}
\hline \hline
\noalign{\smallskip}
  {T$_{\rm sp}$} & {SB} &
 {$B-V$} & {$V-R$} & {$P_{\rm orb}^{1}$} & {$P_{\rm phot}$} &
{\it v}sin{\it i}$^{1}$ &
$\pi$ & $\mu$$_{\alpha}$ cos $\delta$ & $\mu$$_{\delta}$ \\
    &    &      &   & \scriptsize (days) & \scriptsize (days) &
\scriptsize (km s$^{-1}$) &
(mas) & (mas/yr) & (mas/yr) \\
\noalign{\smallskip}
\hline
\noalign{\smallskip}
K0V/K0IV$^{1}$ & 2 & 0.97 & 0.75 & 3.274$^{1}$ & 3.27 & 33.57$^{1}$$\pm$0.45/32
.38$^{1}$$\pm$0.75
 & 9.57$\pm$0.92 & -21.20$\pm$1.30 & -23.00$\pm$1.60\\
\noalign{\smallskip}
\hline
\noalign{\smallskip}
\end{tabular}
\\
$^{1}$ values determined in this paper
\end{flushleft}
\end{table*}

\section{Stellar Parameters of the binary system}
We give the adopted stellar parameters of FF UMa in
Table~\ref{tab:par}. The photometric data ($B-V$, $V$, $P_{\rm phot}$) are
taken from SIMBAD, Jeffries et al. (1995), Henry et al. (1995) and
Strassmeier et al. (2000).
 Orbital period ($P_{\rm orb}$) and projected rotational
velocity ($v\sin{i}$) have been determined in this paper (see
Sects. 3.4 and 4.1). The astrometric data (parallax, $\pi$; proper motions,
$\mu$$_{\alpha}$$\cos{\delta}$ and  $\mu$$_{\delta}$) are from
Hipparcos (ESA 1997) and Tycho-2 (H$\o$g et al. 2000) catalogues.

\subsection{Spectral types and other derived parameters}
To obtain the spectral type of this binary system we
 compared our high resolution echelle spectra, in several
spectral orders free of lines sensitive to chromospheric activity,
with spectra of inactive reference stars of different spectral
types and luminosity classes observed during the same observing
run.
This analysis makes use of the program {\sc starmod} developed
at Penn State University (Barden 1985) which we modified later.
This program constructs a synthesized stellar spectrum
from artificially rotationally broadened, radial-velocity shifted,
and weighted spectra of appropriate reference stars.

For FF UMa we obtained the best fit between observed and synthetic
 spectra using a K1IV reference star
 for primary component and a K0V for the secondary,
 with a contribution to the continuum of 0.70 and 0.30 respectively.
 These spectral types are in agreement with the results reported by
 other authors who suggested an evolved component. In our spectra,
  the spectral features indicate strongly the subgiant nature of the primary.

 We note that since the primary component is a subgiant star,
 the stellar parameters, such us mass and radius, cover a wide range of
 values. Therefore, we determined these crucial characteristics using data
 from the secondary star. Assuming a K0V
spectral type for secondary component, we adopted from
Landolt-B\"{o}rnstein tables (Schmidt-Kaler 1982) a mass $M_{\rm S}$ =
0.79 $M_{\odot}$ and, according to the mass ratio from the orbital
solution ($q$ = 2.12), we derived a primary mass of $M_{\rm P}$ = 1.67
$M_{\odot}$.

In addition, from the photometric period (3.27 days) given by Henry et al.
(1995) and the rotational velocity,
 calculated here, $v\sin{i}_{\rm P}=33.57$ km~s$^{-1}$ (Sect. 4.2), we
 estimated a minimum radius of $R\sin{i}_{\rm P}=2.17$ $R_{\odot}$. This
 agrees with the subgiant radii and previous estimations.

\subsection{Rotational velocities}

Jeffries et al. (1995) estimated the projected rotational velocity
 ($v\sin{i}$) as 41 km s$^{-1}$ for the primary component
 and as 32 km~s$^{-1}$ for the secondary.
Fekel (1997) obtained 38.8 and 39.7 km~s$^{-1}$ for each component
 and Strassmeier et al. (2000) reported lower
 values, 17 and 16 km~s$^{-1}$ respectively.

By using the program {\sc starmod} we obtained the best fits for
each observing run using $v\sin{i}$ values os $\approx$35 km~s$^{-1}$ for
primary component and $\approx$38 km~s$^{-1}$ for secondary component.

To determine a more
accurate rotational velocity of this star we made use of
the cross-correlation technique in our high resolution echelle
spectra by using the routine {\sc fxcor} in IRAF.
The method is described carefully in previous papers (see G\'alvez et al.
 2002; and L\'opez-Santiago et al. 2003) and
 is based on the fact that when a stellar
spectrum with rotationally broadened lines is cross-correlated
against a narrow-lined spectrum, the width of the
cross-correlation function (CCF) is sensitive to the amount of
rotational broadening of the first spectrum.

As a template star in this process we used the K1V star
 \object{HD~26965} for the primary component and the
 K0V star \object{HD~3651} for the secondary for the McDonald run
 and the K2V star \object{HD~166620} for both components in the
 remaining runs.
  All these stars have very low
 rotation velocity -less than 3 km~s$^{-1}$.
 The averaged
 values obtained are  $v\sin{i}$ = 33.57$\pm$0.45 and 32.38$\pm$0.75
 km~s$^{-1}$ for primary and secondary components respectively.

\subsection{Kinematics}

 Computing the galactic space-velocity components ($U$, $V$, $W$) of
FF~UMa required both radial velocity and precise proper motions
 and parallax. For the former, we used the center of mass velocity, $\gamma$,
 determined in the orbital solution for the FOCES04 observing run (see
Sect.~3.4). For the latter we utilized data taken from
Hipparcos (ESA 1997) and Tycho-2 (H$\o$g et al. 2000) catalogues
 (see Table~\ref{tab:par}) (for details see Montes et al. 2001a, b).

In addition, we included FF~UMa in an extended study of binary star
 kinematics in young moving groups. It included the application of
 Eggen's peculiar velocity and radial velocity criteria
 (see Montes et al. 2001a and reference therein)
  and spectroscopic criteria (see the
  Li~{\sc i} $\lambda$6707.8 line Sect. 4.4).

The resulting values of ($U$, $V$, $W$) and associated errors are given in
Table~\ref{tab:uvw}. These errors have been calculated assuming the
value of $\gamma$ determined from the FOCES04 observing run. However,
 taking into account the changes in $\gamma$ $\approx$ 3 km~s$^{-1}$ we
 detected, we expect the uncertainties to be larger (see Sect. 3.3).

Using the ($U$, $V$) and ($V$, $W$) diagrams
 (Eggen 1984, 1989; Montes et al. 2001a), the velocity components lie
 clearly inside the Castor moving group boundaries. In addition,
 Eggen's radial velocity criteria also confirm their membership
 of FF UMa to this group (G\'alvez 2005).

\begin{table}
\caption[]{Galactic space-velocity components
\label{tab:uvw}}
\begin{flushleft}
\small
\begin{center}
\begin{tabular}{cccc}
\noalign{\smallskip}
\hline \hline
\noalign{\smallskip}
$U\pm \sigma_{U}$ & $V \pm \sigma_{V}$ & $W \pm \sigma_{W}$ & $V_{\rm Total}$ \\
(km s$^{-1}$) & (km s$^{-1}$) & (km s$^{-1}$)  & (km s$^{-1}$) \\
\noalign{\smallskip}
\hline
\noalign{\smallskip}
-7.04$\pm$1.14 & -12.92$\pm$1.33 & -6.68$\pm$0.68 & 16.16 \\
\noalign{\smallskip}
\hline
\noalign{\smallskip}
\end{tabular}
\end{center}
\end{flushleft}
\end{table}

\subsection{The Li~{\sc i} $\lambda$6707.8 line}

As it is well known, Li~{\sc i} $\lambda$6707.8 spectroscopic feature
is an important diagnostic of age in late-type stars,
 since it is destroyed easily by thermonuclear reactions
in the stellar interior.

The spectral region of the resonance doublet of  Li~{\sc i} at
$\lambda$6708 \AA\ is covered by most of our spectral observations.
 Despite blending with photospheric lines, mainly Fe~{\sc i} (6707.4 \AA),
we could separate the contribution from both components. We then measured
 the equivalent width ($EW$ hereafter) of (Li~{\sc i} + Fe~{\sc i})
 of both components in our
 observed spectra. We calculated the contribution of Fe~{\sc i} by
 using both calibrations of Fe~{\sc i}-effective temperature from
 Soderblom et al. (1990) and Fe~{\sc i}-($B-V$) color index from
 Favata et al. (1993). We obtained the corrected
 $EW$(Li~{\sc i}) by subtracting
 the $EW$(Fe~{\sc i}) of the total measured equivalent width,
 $EW$(Li~{\sc i}+Fe~{\sc i}). The resulting mean values of
 $EW$(Li~{\sc i}) are 200 m\AA\ for the primary
 component and 141 m\AA\ for the secondary. These values are corrected by the
 contribution of each component to the continuum (see Sect. 5).

 By using the spectral subtraction technique, that is, obtaining the
 $EW$ of Li I directly from the subtracted spectra, we obtained a
   mean $EW$(Li~{\sc i}) of 132 m\AA\
 and 86 m\AA\ for both component respectively.

 The discrepancy between the two methods used to calculate the $EW$
 is due to the influence of stellar metallicities. In the first technique,
 stellar metallicity is not taken into account in the relation
 calibrations, and in the second technique, the stellar metallicity
 of the standard star used to create the synthetic spectra is undetermined.
 In spite of this, these $EW$(Li~{\sc i})s values are of the same order as
  the Li~{\sc i} $EW$s of other Castor moving group members, which have
 an age around 200 Myr.

\begin{figure}
{\psfig{figure=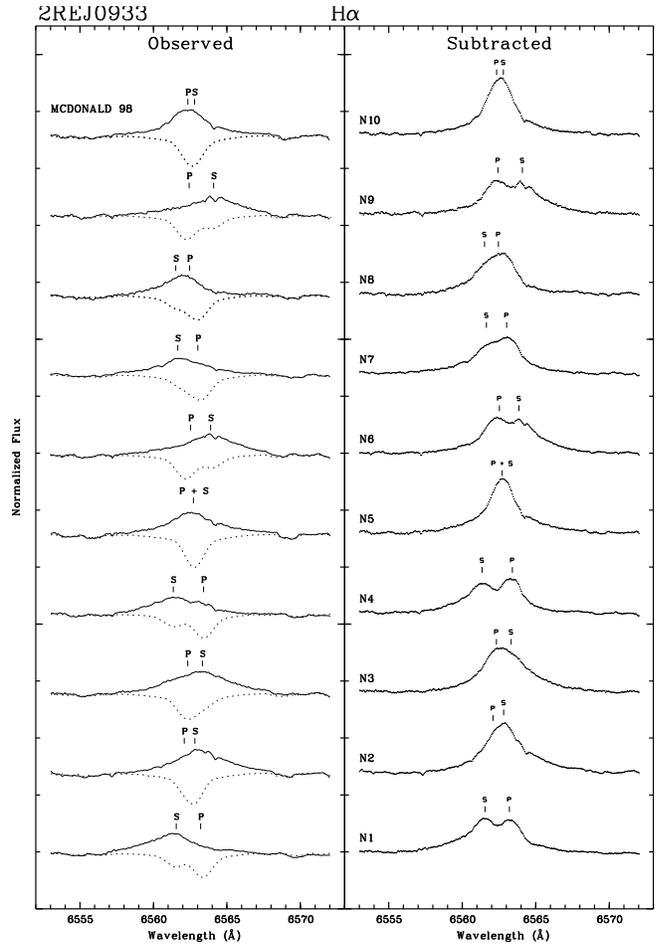,bbllx=34pt,bblly=34pt,bburx=545pt,bbury=774pt,height=12.5cm,width=8.6cm,clip=}}
\caption[]{Spectra in the H$\alpha$ line region in McDonald98 observing
run.
 We plot the observed spectrum (solid-line) and the synthesized
 spectrum (dashed-line) in the left panel, and the subtracted spectrum
 (dotted line), in the right panel. We mark position of primary component
 lines  with a (P) and position of secondary component lines with a (S).
\label{fig:ha}}
\end{figure}

\section{Chromospheric activity indicators}
The echelle spectra analyzed allow us to study the behavior of
 the Ca~{\sc ii} H \& K to the Ca~{\sc ii} IRT lines,
 different indicators formed at varying atmospheric heights. We
 determined the chromospheric contribution of these features
 using the spectral subtraction technique described in
detail by Montes et al. (1995) and Papers~I, II, III and IV. We
constructed the synthesized spectrum using the program {\sc starmod}.

Taking into account the stellar parameters derived in Sect.~4 we
 used reference stars of the K1IV spectral type for the primary component and
K0V spectral type for the secondary component (see Sect.~4.3), with a
 contribution of 0.70/0.30 respectively.

In Table~\ref{tab:ew} (only available in electronic form) we present the
excess emission $EW$, measured in the
 subtracted spectra, for the Ca~{\sc ii} H \& K, H$\epsilon$,
H$\delta$, H$\gamma$, H$\beta$, H$\alpha$, and  Ca~{\sc ii} IRT
 lines in all observing runs. We list
 the $EW$s of emission features for both components (P/S);
 when lines were blended we list only total $EW$s.
The uncertainties in the measured $EW$ were estimated taking into account:
a) the typical internal precisions of {\sc starmod}
(0.5 - 2 km s$^{-1}$ in velocity shifts, and $\pm$5 km s$^{-1}$ in  $v\sin{i}$),
b) the rms obtained in the fit between observed and
synthesized spectra in the spectral regions outside the chromospheric features
(typically in the range 0.01-0.03) and
c) the standard deviations of the $EW$ measurements.
The final estimated errors are in the range 10-20\%.

We corrected the measured $EW$s for
the relative contribution of each component to the total continuum
 ($S_{\rm P}$ and $S_{\rm S}$), using the radii assumed in Sect.~4.1.
 and temperatures
  from Landolt-B\"{o}rnstein tables (Schmidt-Kaler 1982).
We obtained the final $EW$s for the components
  multiplying by a factor $1/S_{\rm P}$ and $1/S_{\rm S}$, respectively.
We present the result in Table~\ref{tab:ew}.

These adopted $EW$s were transformed to absolute
 surface fluxes using the
empirical stellar flux scales calibrated by Hall (1996) as a function of
the star color index.
 We used the $B-V$ index and the corresponding coefficients
for Ca~{\sc ii} H \& K, H$\alpha$ and Ca~{\sc ii} IRT. We used for H$\epsilon$
the same coefficients as for Ca~{\sc ii} H \& K, and derived the H$\delta$,
 H$\gamma$ and H$\beta$ coefficients of flux by making an interpolation
 between the values of Ca~{\sc ii} H \& K and H$\alpha$.
We present the logarithm of the obtained absolute flux at the stellar surface
(log$F_{\rm S}$) for the different chromospheric activity indicators
 in Table~\ref{tab:flux} (only available in electronic form).

\begin{center}

\begin{figure}
{\psfig{figure=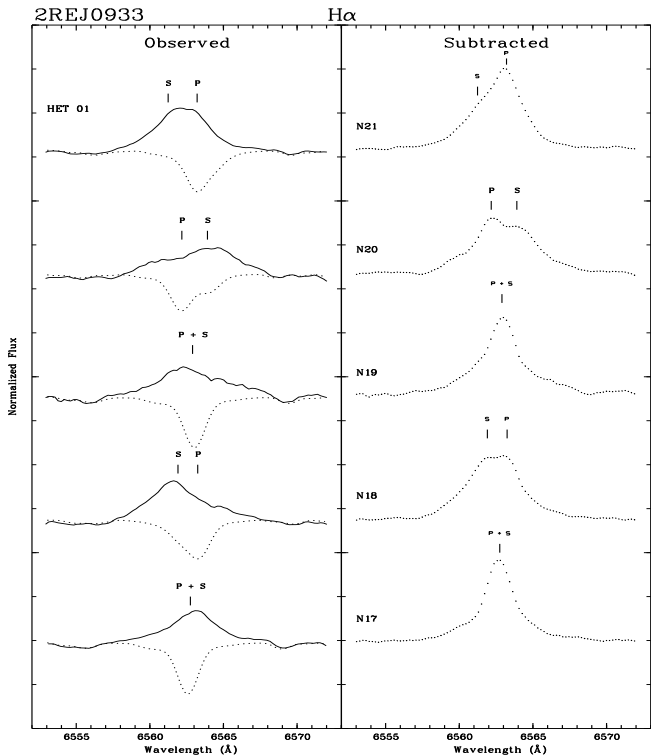,bbllx=34pt,bblly=32pt,bburx=544pt,bbury=775pt,height=10.0cm,width=8.6cm,clip=}}
\caption[] {The same as in previous Figure in HET00 observing run.
\label{fig:ha2}}
\end{figure}
\end{center}

\begin{center}
\begin{figure}
\psfig{figure=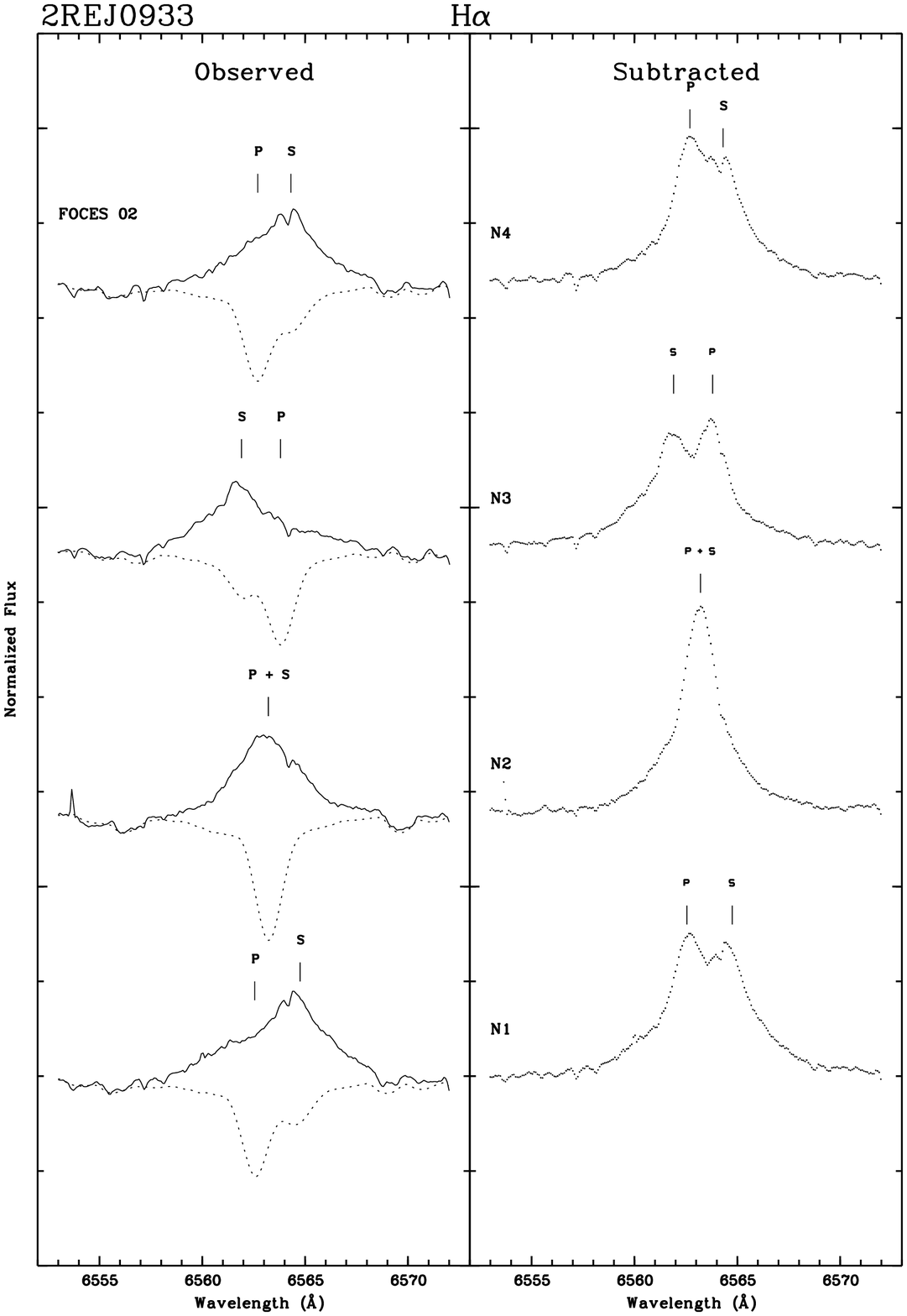,bbllx=34pt,bblly=34pt,bburx=544pt,bbury=775pt,height=10.0cm,width=8.6cm,clip=}
\caption[] {The same as in previous Figure in FOCES02 observing run.
\label{fig:ha3}}
\end{figure}
\end{center}

\begin{figure}
\psfig{figure=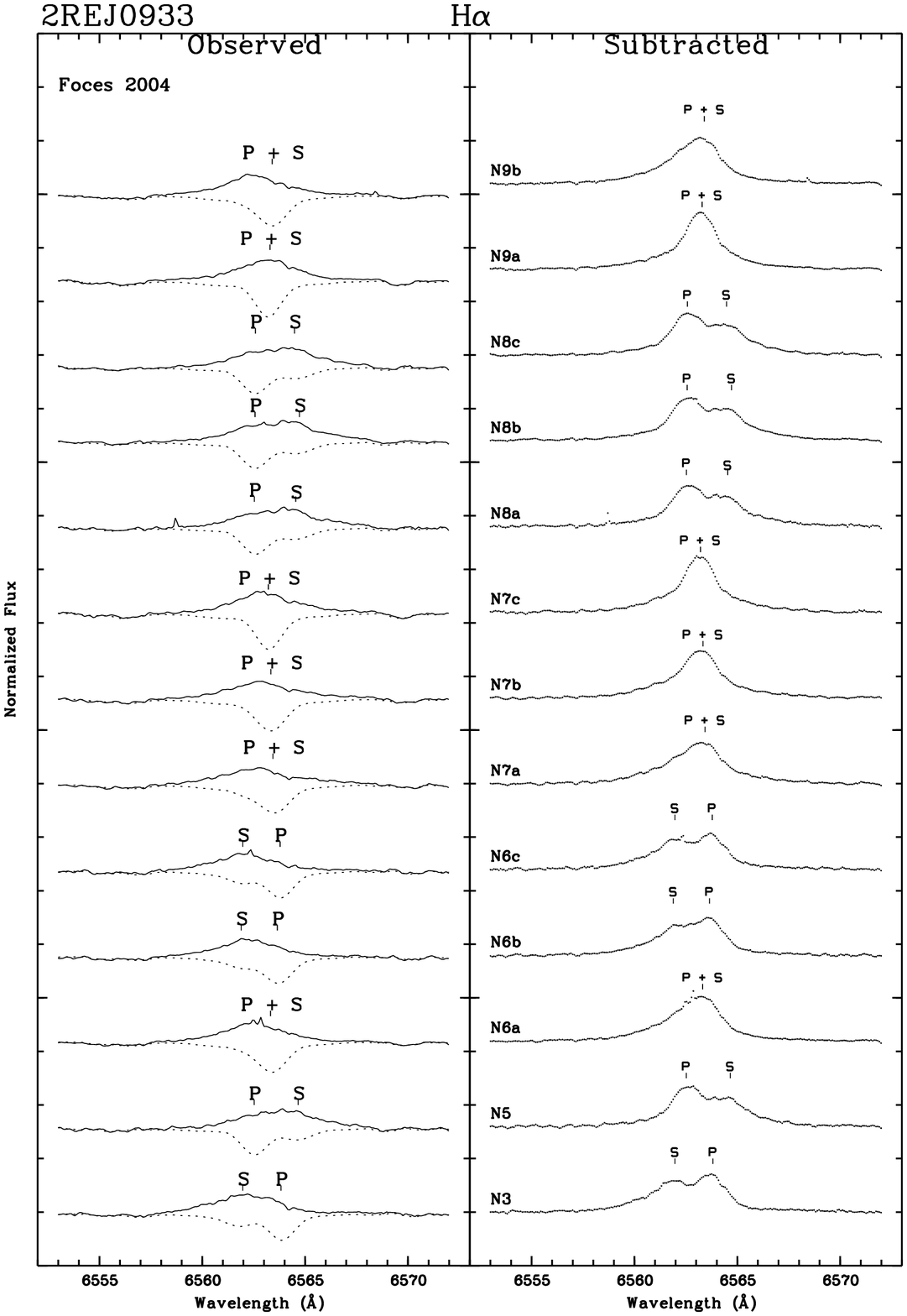,bbllx=34pt,bblly=34pt,bburx=544pt,bbury=775pt,height=12.5cm,width=8.6cm,clip=}
\caption[] {The same as in previous Figure in FOCES04 observing run.
\label{fig:ha4}}
\end{figure}

\begin{center}
\begin{figure}
\psfig{figure=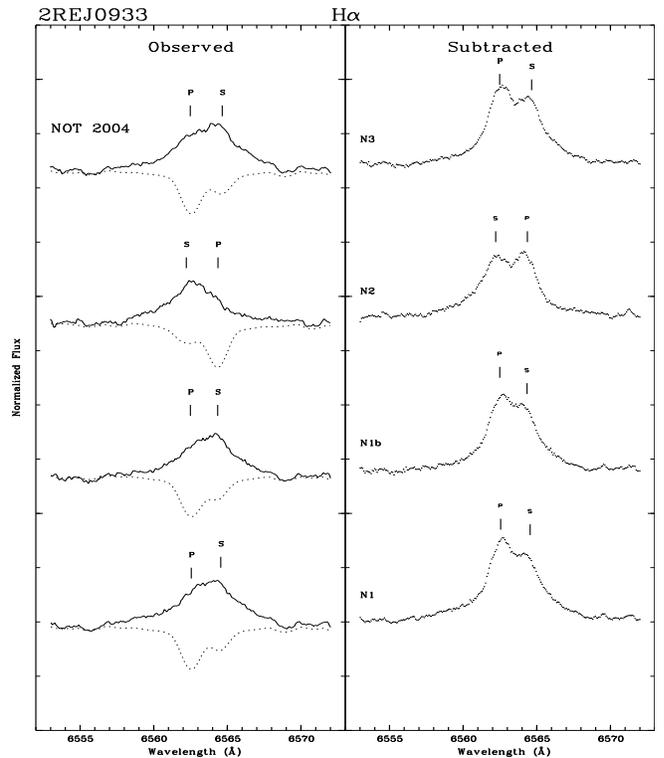,bbllx=34pt,bblly=34pt,bburx=544pt,bbury=775pt,height=10.0cm,width=8.6cm,clip=}
\caption[] {The same as in previous Figure in NOT04 observing run.
\label{fig:ha5}}
\end{figure}
\end{center}

In Figs.~\ref{fig:ha} -~\ref{fig:irt3} we plot
 the H$\alpha$, Ca~{\sc ii} H \& K  and
Ca~{\sc ii} IRT $\lambda$$8498$, $\lambda$$8542$ lines region
 for each observation, the observed spectrum (solid-line)
and the synthesized spectrum (dashed-line) in the left panel
and the subtracted spectrum (dotted line) in the right panel.
We included the observing run of each spectrum in these figures.
In Fig.~\ref{fig:hb}, we plot a representative subtracted spectrum
 of FF~UMa in the H$\beta$ line region.


\subsection{H$\alpha$}

The H$\alpha$ line is observed in emission above the continuum
 in all the spectra
(see Figs.~\ref{fig:ha} -~\ref{fig:ha5}, left panel).
In the observed spectrum, the emission associated with the secondary is
 larger than that associated with the primary. However, after applying
 the spectral subtraction
technique, the H$\alpha$ emissions above the continuum coming from both
components are similar and in some cases the primary one is larger.
The H$\alpha$ emission is persistent during all observations
 indicating that it is
a very active binary system similar to RS~CVn and BY~Dra systems
 that always show H$\alpha$ emission above the continuum.
Measuring the $EW$ of this line, we found that each stellar component is
 formed by a central narrow component and a broad component that moves
 from red to blue. These are an indication of microflare activity
 (see Papers I, II, and III).
While we were able to separate the narrow components, we were unable to
 deblend the broad ones. Therefore, to determine the contribution of each
stellar component to the total excess emission, we fit the
narrow and broad component of each star together
 (see Figs.~\ref{fig:lorent} and~\ref{fig:lorent2}).

The $EW$ average value measured in the subtracted spectra is
  $EW$(H$\alpha$)~=~1.64/2.47 \AA\ for the primary and secondary
 components. We note that these are higher values than those reported
 by Jeffries et al. (1995).
In Table~\ref{tab:ew} we list the $EW$ of each stellar component
 determined by the fit
 described above. We also list the total $EW$ (primary $+$ secondary)
 determined by integrating the total excess emission profile.
We note that H$\alpha$ line shows notable variations with orbital phase
 but also from one epoch to another in both components.

\subsection{H$\beta$, H$\gamma$ and H$\delta$}

We can see absorption of H$\beta$, H$\gamma$ and H$\delta$ Balmer
 lines filled in with emission in the observed spectra.
 After applying the spectral
 subtraction, clear excess emission is detected from both components (see
 a representative spectrum in the H$\beta$ line region in
 Fig.~\ref{fig:hb}). When the $S/N$ was high enough we deblended the
  emission coming from both components by using a two-Gaussian fit to
 the subtracted spectra (see Table~\ref{tab:ew}).
These three lines show the same behavior with orbital phase that
 the H$\alpha$ line in both components. Their mean values are
$EW$(H$\beta$)~=~0.33/0.25 \AA, $EW$(H$\gamma$)~=~0.14/0.19 \AA\
and $EW$(H$\delta$)~=~0.13/0.15 \AA.

We also measured the ratio of
excess emission in the H$\alpha$ and H$\beta$ lines,
$\frac{EW({\rm H\alpha})}{EW({\rm H\beta})}$, and the ratio
of excess emission $\frac{E_{\rm H\alpha}}{E_{\rm H\beta}}$
with the correction:

\[ \frac{E_{\rm H\alpha}}{E_{\rm H\beta}} =
\frac{EW({\rm H\alpha})}{EW({\rm H\beta})}*0.2444*2.512^{(B-R)}\]
given by Hall \& Ramsey (1992). This takes into account
the absolute flux density in these lines and the color difference
in the components.
We obtained mean values of
 $\frac{E_{\rm H\alpha}}{E_{\rm H\beta}}$ $\approx6$ for the primary
 component and $\approx5$ for the secondary.
 These values indicate, according to Buzasi (1989) and Hall \& Ramsey (1992),
 the presence of prominence-like material above the stellar
 surface in both components of the system.

\begin{figure}
{\psfig{figure=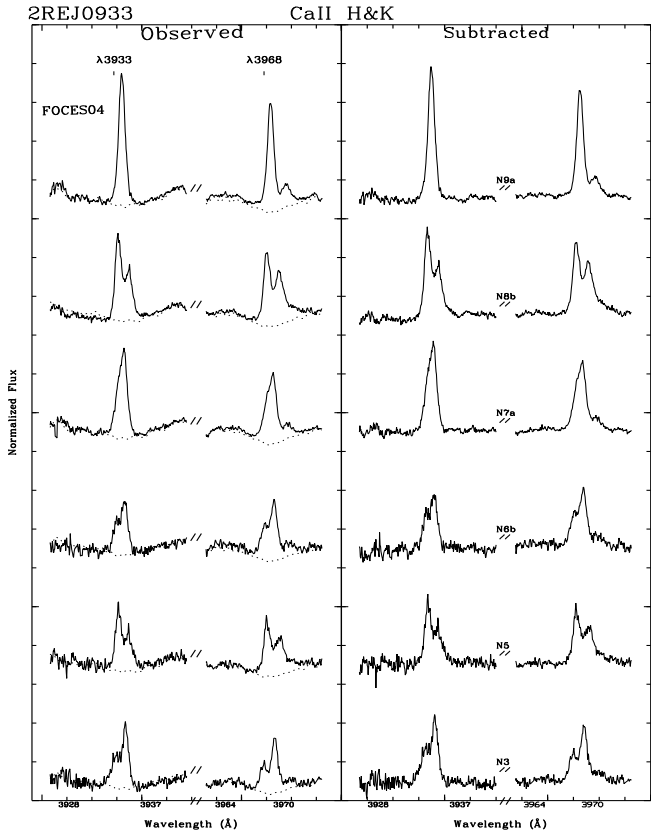,bbllx=34pt,bblly=34pt,bburx=544pt,bbury=775pt,height=11.0cm,width=8.6cm,clip=}}
\caption[] {The same as previous Figure in Ca~{\sc ii} H \& K line region
in FOCES04 observing run. \label{fig:hyk}}
\end{figure}

\subsection{Ca~{\sc ii} H \& K and H$\epsilon$}
The Ca~{\sc ii} H \& K line region is included in
FOCES 2002 and 2004 and NOT04 observing runs.

This spectral region is located at the end of the echellogram,
 where the efficiency of the spectrograph and the CCD decrease
 very rapidly and therefore the
 $S/N$ ratio obtained is very low; thus the
  normalization of the spectra
is very difficult. In spite of this, the spectra show strong emission
in the Ca~{\sc ii} H $\&$ K lines and a clear emission in the H$\epsilon$ line
 from both components (see Fig.~\ref{fig:hyk}). These
 allow us to apply the spectral subtraction in this spectral
 region. As we can see in Fig.~\ref{fig:hyk}, the
 H$\epsilon$ line arising from one of the component overlaps with
 the Ca~{\sc ii} H line arising from the other component at some orbital
  phases, so their $EW$ were measured with a
 Gaussian fit when it was possible.
Mean $EW$s values measured in these spectra are
 $EW$=~1.23/1.14 \AA\ for each component in Ca~{\sc ii} K line,
$EW$=~1.23/1.08 \AA\ in Ca~{\sc ii} H line and
 $EW$=~0.30/0.43 \AA\ in H$\epsilon$ line.

As the H$\alpha$ emission line, the Ca~{\sc ii}  H \& K lines show
 variations with both orbital phase and from one epoch to another
 in both components.

\begin{figure}
\begin{center}
{\psfig{figure=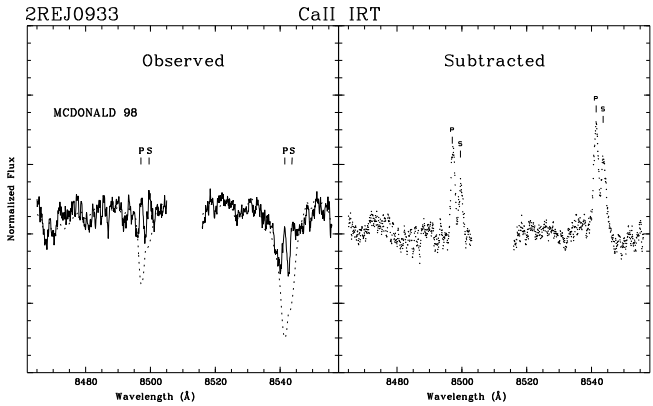,bbllx=195pt,bblly=33pt,bburx=559pt,bbury=629pt,height=5.3cm,width=8.6cm,angle=270,clip=}}
\caption[] {The same as in previous figures but in Ca~{\sc ii} IRT
($\lambda$8498 \& $\lambda$8542)
 line regions in McDonald98 observing run.
\label{fig:irt}}
\end{center}
\end{figure}
\begin{figure}
{\psfig{figure=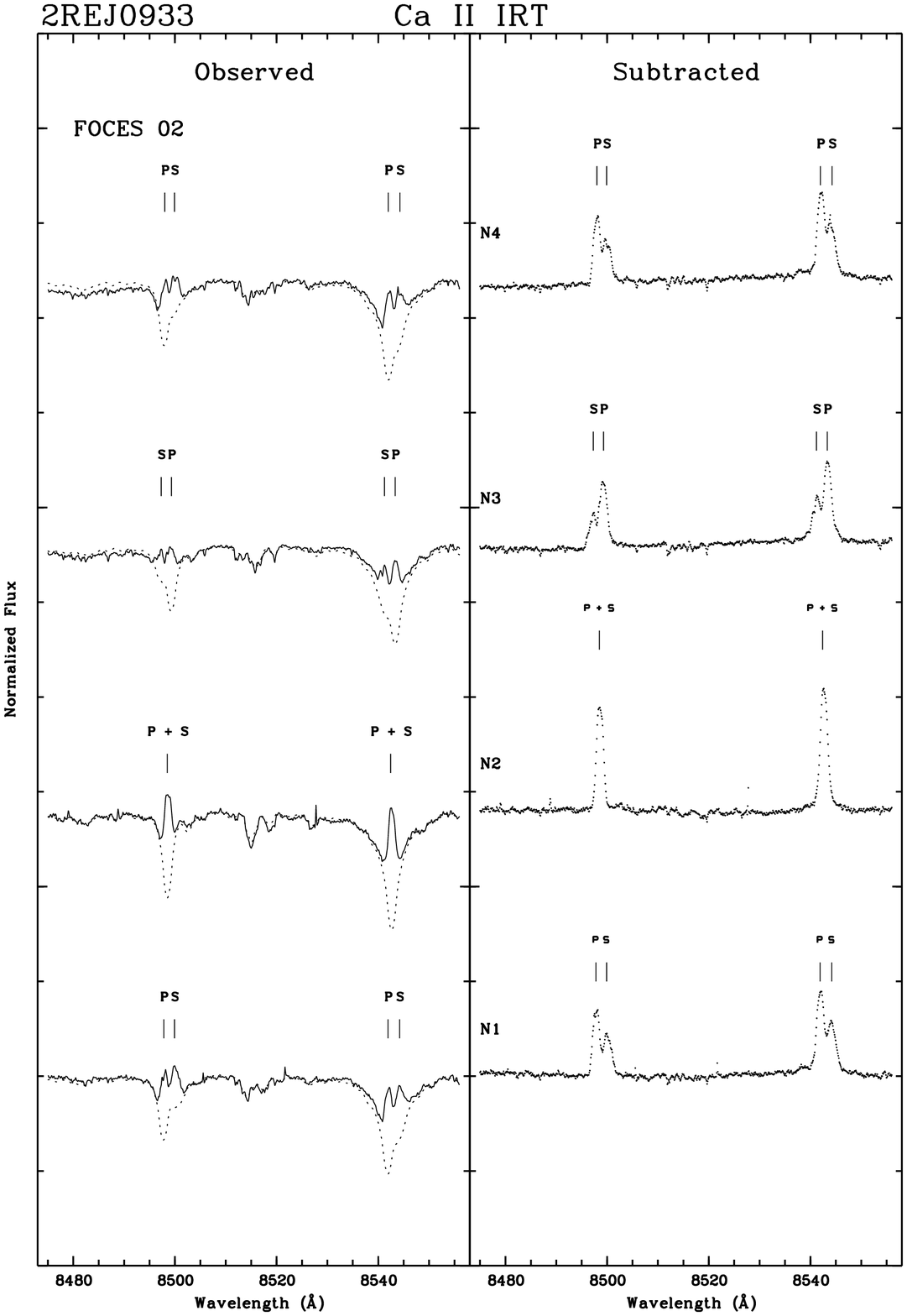,bbllx=34pt,bblly=34pt,bburx=545pt,bbury=775pt,height=10.0cm,width=8.6cm,clip=}}
\caption[] {The same as previous Figure in FOCES02 observing run.
\label{fig:irt2}}
\end{figure}
\begin{figure}
{\psfig{figure=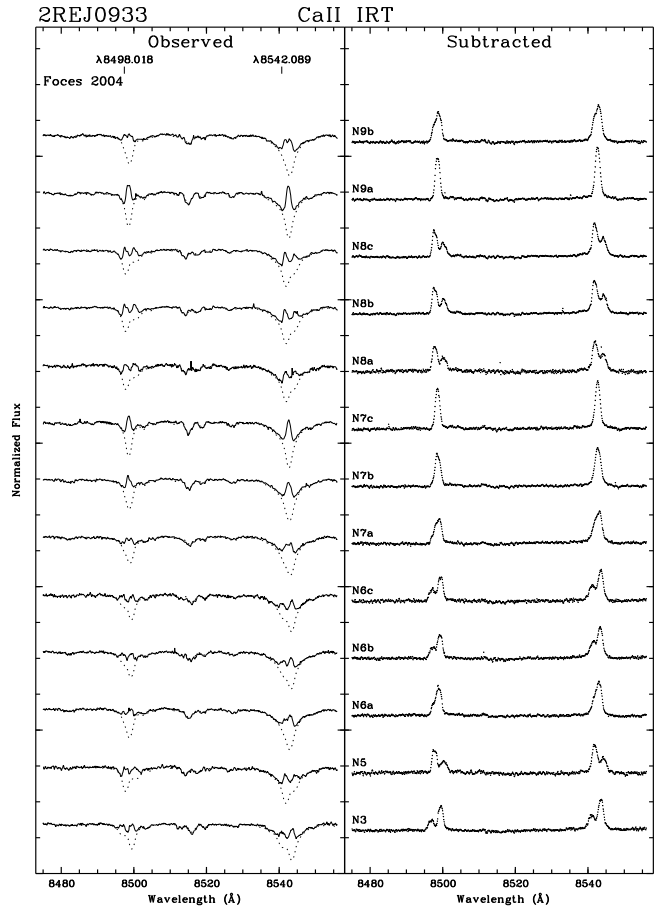,bbllx=34pt,bblly=34pt,bburx=545pt,bbury=775pt,height=12.0cm,width=8.6cm,clip=}}
\caption[] {The same as previous Figure in FOCES04 observing run.
\label{fig:irt3}}
\end{figure}

\subsection{Ca~{\sc ii} IRT lines
($\lambda$8498, $\lambda$8542 and $\lambda$8662)}

All our echelle spectra include the three lines of the Ca~{\sc ii}
 infrared triplet (IRT) except for the $\lambda$8498 line in
HET00 and NOT04 runs.
In all of the spectra we observed a clear emission above the continuum
 in the core of the Ca {\sc ii} IRT absorption lines
(see Figs.~\ref{fig:irt} -~\ref{fig:irt3}) from both components.
After applying the spectral subtraction, we could see
that the emission coming from the primary component is larger than
the emission from the secondary.

We measured mean $EW$s for these three Ca {\sc ii} lines of
$EW$($\lambda$8498, $\lambda$8542, $\lambda$8662)=~0.58/0.35,~0.67/0.43
 and~0.59/0.35 \AA \ respectively.
For each component we found considerable variations with orbital
phase that appear anti-correlated with the variations in the
 Balmer lines.

In addition, we calculated the ratio of excess emission $EW$,
$\frac{E_{8542}}{E_{8498}}$, which is also an indicator of the
type of chromospheric structure that produces the observed emission;
 in solar plages, values of
 $\frac{E_{8542}}{E_{8498}}$ $\approx$~1.5-3 are measured,
 while in solar prominences the values are $\approx$~9,
 the limit of an optically thin emitting plasma (Chester 1991).
 We found for this star small values of the $\frac{E_{8542}}{E_{8498}}$
 ratio, $\approx$~1.0, for both components
(see Table~~\ref{tab:flux}). This indicates
that the Ca~{\sc ii} IRT emission of this star arises from
plage-like regions at the stellar surface,
 in contrast with the Balmer lines that come from prominences.
This markedly different behavior of the Ca~{\sc ii} IRT emission
has also been found in other chromospherically active binaries (see
Papers~III; IV and references therein).

\begin{figure}
{\psfig{figure=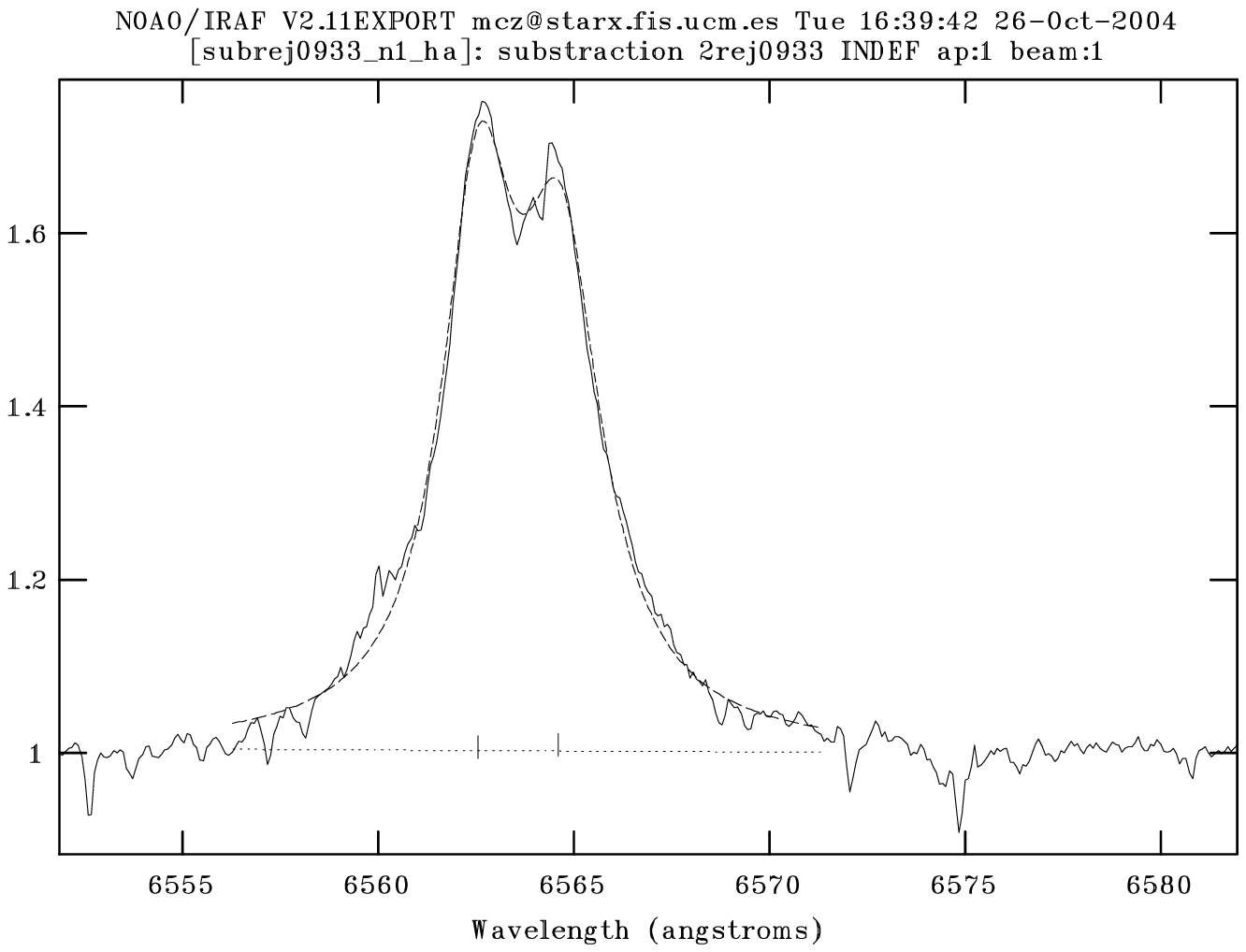,bbllx=72pt,bblly=242pt,bburx=540pt,bbury=528pt,height=6.0cm,width=8.3cm,clip=}}
\caption[] {Example of the H$\alpha$ region fit in the subtracted spectrum
of the two
 components by using IRAF {\it SPLOT} task.
\label{fig:lorent}}
\end{figure}
\begin{figure}
{\psfig{figure=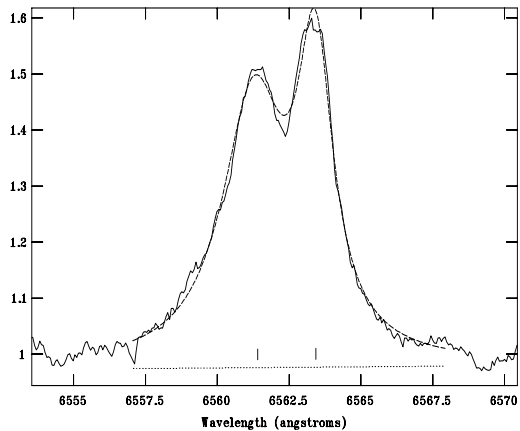,bbllx=72pt,bblly=242pt,bburx=540pt,bbury=528pt,height=6.0cm,width=8.3cm,clip=}}
\caption[]{Another example of H$\alpha$ region fit. \label{fig:lorent2}}
\end{figure}

\begin{figure}
{\psfig{figure=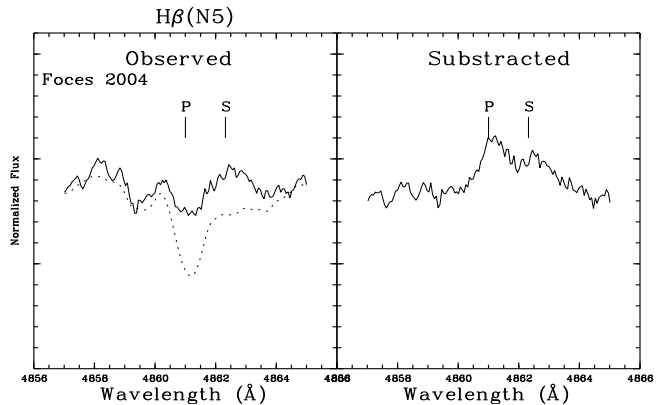,bbllx=221pt,bblly=28pt,bburx=563pt,bbury=540pt,height=5.3cm,width=8.6cm,angle=270,clip=}}
\caption[] {The same as H$\alpha$ figures but in H$\beta$
 line region in FOCES04 observing run.
\label{fig:hb}}
\end{figure}

\subsection{Variation of activity with time}
As we mentioned above, the $EW$ emission lines that are chromospheric
 indicators show variations with orbital phase due to activity features
 present in both stellar surfaces.
 But there are also variations from one epoch to another in both components.
 To study if there is a correlation between active cycle and orbital
 period variation as Lanza (2006) suggested for the case of HR~1099, we
 must have a follow up of this system during a complete cycle of its
 variation period ($\geq$ 22 years). This kind of study would
 provide us another clue for understanding and testing
 the Applegate's mechanism.

\section{Conclusions}

In this paper, we present a detailed spectroscopic analysis of a
 X-ray/EUV selected chromospherically active binary system
 2RE~J0933+624 (FF~UMa). We analyzed high resolution
 echelle spectra that include
the optical chromospheric activity indicators from the Ca~{\sc ii} H \& K
to Ca~{\sc ii} IRT lines, as well as the
 Li~{\sc i} $\lambda$6707.8 line and other photospheric lines of interest.

With a large number of radial velocities from the literature and
from our spectra taken over several years, we found
 that this system shows an orbital period variation similar to those
 previously found in other RS~CVn systems.

 Although the existence of an unseen distant third star as a component
  of the system cannot be completely ruled out as the cause of these
 variations, we think that the more plausible explanation is
  Applegate's mechanism, or at least the qualitative idea; which states that
 the orbital period change is due to the gravitational coupling
 of the orbit to changes in the quadrupole moment of the magnetically
 active stellar components of the system. In the case of FF~UMa, we
 calculated a ($O-C$) ($T_{\rm conj}$) that gives us a
 relative orbital period variation of
  $dP/P$ $\approx$ 10$^{-4}$ in 11 years, that is, one order
 of magnitude higher than the variations in HR~1099, the largest
 observed until now.
 We suggest here that this order of magnitude difference between
 the period variations in FF~UMa and HR~1099 could be explained
 by the different activity level.
 The components of FF~UMa are very active and have more effective
 dynamo mechanisms than HR~1099 components.


 Once we adopted an orbital period, from
 the FOCES04 observing run, we improved
 the determination of the orbital solution of the system
 relative to previous determinations by other authors.
 We obtained a nearly circular orbit with an orbital period
 very close to photometric period,
 indicating that it has a synchronous rotation.

 The spectral classifications derived by comparing FF UMa with spectra of
reference stars, leads us to consider the primary component as a
 subgiant star and the secondary component as a K0V star. The results
 from orbital parameters and photometric characteristics
 help us to obtain physical parameters from the primary,
 $M_{\rm P}$ = 1.67 $M_{\odot}$ and $R\sin{i}_{\rm P}=2.17$ $R_{\odot}$, in
 agreement with subgiant radii and previous estimates.

By using the information provided by the width of the cross-correlation
function we determined a projected rotational
velocity, $v\sin{i}$, of 33.57$\pm$0.45 km s$^{-1}$ and
 32.38$\pm$0.75 km~s$^{-1}$ for the primary and secondary
components respectively.

The presence of the Li~{\sc i} line is in agreement
with the kinematics results, i.e., it belongs to the young disk
 and is probably a member of the Castor moving group.

The study of the optical chromospheric activity indicators
 shows that FF~UMa system has a high level of activity in both components.
The variation of H$\alpha$ and the rest of the Balmer and Ca~{\sc ii}
H$\&$K lines are very similar and anti-correlated in phase with Ca~{\sc ii} IRT
 emission, as we confirmed with the
 $\frac{E_{\rm H\alpha}}{E_{\rm H\beta}}$ and
 $\frac{E_{8542}}{E_{8498}}$ results. This indicates that the Balmer
 emission lines arise from prominence-like material while the
 emission of Ca~{\sc ii} IRT lines arise from plage-like regions.
In addition, both components show variations from one epoch to another
 that could have a correlation with the orbital period variation.
 Future spectroscopic and photometric studies of this system
 could confirm this hypothesis and provide a better understanding and test
 of Applegate's mechanism.



\begin{acknowledgements}
We would like to thank Dr. L.W. Ramsey for collaborating on the
McDonald observing runs (2.1m and HET telescopes) and the staff of
McDonald observatory for their allocation of observing time and
their assistance with our observations. We thank Michelle L. Edwards
 for her help in checking and correcting the English writing.
 Finally, we want to thank the anonymous referee for very useful comments.
This work was supported by the Universidad Complutense de Madrid and the
Spanish Ministerio Educaci\'on y Ciencia (MEC), Programa Nacional de
Astronom\'{\i}a y Astrof\'{\i}sica under grant AYA2005 - 02750
  and the "Comunidad de Madrid" under
PRICIT project S-0505/ESP-0237 (ASTROCAM).
\end{acknowledgements}



\onltab{7}{
\begin{table*}
\caption[]{$EW$ of chromospheric activity indicators \label{tab:ew}}
\begin{flushleft}
\scriptsize
\begin{tabular}{lccccccccccc}
\noalign{\smallskip}
\hline \hline
\noalign{\smallskip}
 &          & \multicolumn{10}{c}{$EW$(\AA)  in the subtracted spectra} \\
\cline{3-12} \noalign{\smallskip} Obs. &  $\varphi$ &
\multicolumn{2}{c}{Ca {\sc ii}}  & & & & & &
\multicolumn{3}{c}{Ca {\sc ii} IRT} \\
\cline{3-4}\cline{10-12} \noalign{\smallskip} Idt.$^{5}$  &   &
 K   & H  & H$\epsilon$ & H$\delta$ & H$\gamma$ & H$\beta$ & H$\alpha$ &
    $\lambda$8498 &      $\lambda$8542 &     $\lambda$8662
\\
 & & & & &  & &  & H$\alpha$$_{i}$(P+S)& & & \\
\noalign{\smallskip}
\hline
\noalign{\smallskip}
(1) & 0.75 & - & - & - & - & - & - & 0.89/2.45 & 0.35/0.23 & 0.53/0.41 & 0.48/0.28 \\
 & & & & & & & &  2.66  & &\\
(1) & 0.06 & - & - & - & - & - & - & 3.64$^{1}$ & 0.63$^{1}$ & 1.02$^{1}$ & 0.77$^{1}$ \\
 & & & & & & & &  3.27  & &\\
(1) & 0.37 & - & - & - & - & - & - & 1.54/2.93 & 0.58$^{1}$ & 1.18$^{1}$ & 0.55$^{1}$ \\
 & & & & & & & &  3.31  & &\\
(1) & 0.67 & - & - & - & - & - & - & 1.24/2.02 & - & - & - \\
 & & & & & & & &  2.43  & &\\
(1) & 0.98 & - & - & - & - & - & - & 3.52$^{1}$ & 0.54$^{1}$ & 0.88$^{1}$ & 0.66$^{1}$ \\
 & & & & & & & &  2.96  & &\\
(1) & 0.30 & - & - & - & - & - & - & 1.43/2.21 &0.38/0.22 & 0.60/0.43 & 0.48/0.27 \\
 & & & & & & & &  2.88  & &\\
(1) & 0.60 & - & - & - & - & - & - & 0.90/2.31 & 0.43/0.10 & 0.80$^{1}$ & 0.53/0.28 \\
 & & & & & & & &  2.47  & &\\
(1) & 0.90 & - & - & - & - & - & -  & 1.00/2.15 & $^{3}$ & $^{3}$ & $^{3}$ \\
 & & & & & & & &  2.62  & &\\
(1) & 0.21 & - & - & - & - & - & -  & 1.25/2.19 & 0.37/0.23 & 0.70/0.56 & 0.42/0.31 \\
 & & & & & & & &  2.70  & &\\
(1)   & 0.55 & - & - & - & - & - & -  & 1.25/1.93& 0.57$^{1}$ & 0.94$^{1}$ & 0.53$^{1}$ \\
 & & & & & & & &  3.07  & &\\
(2) & 0.74 & - & - & - & - & - & 0.31/0.23 & 1.95/2.73 & - & 0.71/0.32 & 0.58/0.32 \\
 & & & & & & & &  4.90  & &\\
(2) & 0.07 & - & - & - & - & - & 0.59$^{1}$ & 3.42$^{1}$ & - & 1.07$^{1}$ & 0.93$^{1}$ \\
 & & & & & & & &  2.93 & &\\
(2) & 0.47 & - & - & - & - & - &  0.65$^{1}$ & 3.53 $^{1}$ & - & 1.16 $^{1}$ & 1.03$^{1}$ \\
 & & & & & & & &  3.38  & &\\
(2) & 0.30 & - & - & - & - & - & 0.88$^{1}$ & 1.62/3.09  & - & 0.74/0.41 & 0.72/0.14 \\
 & & & & & & & &  3.23  & &\\
(2) & 0.26 & - & - & - & - & - &  1.13$^{1}$ & 5.20 $^{1}$ & - & 1.25 $^{1}$ & 1.14$^{1}$ \\
 & & & & & & & &  4.03  & &\\
(2) & 0.62 & - & - & - & - & - & 0.68$^{1}$ & 3.95$^{1}$ & - & 1.11$^{1}$ & 0.95$^{1}$ \\
 & & & & & & & &  3.33  & &\\
(2) & 0.91 & - & - & - & - & - & 0.66$^{1}$ & 2.58/1.58  & - & 0.68/0.30 & 0.59/0.43 \\
 & & & & & & & &  3.19  & &\\
(2) & 0.26 & - & - & - & - & - & 0.91$^{1}$ & 3.42/1.57 & - & 1.26$^{1}$ & 1.34$^{1}$ \\
 & & & & & & & &  3.45  & &\\
(3)   & 0.56 & 1.80/1.96 & 1.46/2.17 & $^{3}$ & $^{3}$ & 0.07/0.13$^{2}$& 0.46/0.34 &  2.15/1.99 &0.80/0.60 & 0.55/0.43 & 0.67/0.42 \\
 & & & & & & & &  3.70  & &\\
(3) & 0.86 & 3.42$^{1}$ & 2.45$^{1}$ & 0.62$^{1}$ & 0.20$^{1}$ & 0.26$^{1,}$$^{2}$ & 0.74$^{1,}$$^{2}$ & 3.99$^{1}$ & 1.20$^{1}$ & 0.93$^{1}$ & 1.04$^{1}$\\
 & & & & & & & &  3.34  & &\\
(3) & 0.17 & 1.92/1.27 & 1.31/0.98 & 0.589/$^{3}$ & 0.09/0.10 & 0.12/0.15$^{2}$
& 0.45/0.21$^{2}$ & 1.33/2.01 & 0.91/0.38 & 0.63/0.22 & 0.73/0.32\\
 & & & & & & & &  2.91  & &\\
(3) & 0.50 &1.88/1.81 & 1.53/1.11& 0.41/0.07 & 0.10/0.11 & 0.07/0.16$^{2}$ & 0.21/0.45$^{2}$ & 1.85/1.81 & 0.78/0.65 & 0.49/0.39 & 0.73/0.43\\
 & & & & & & & &  3.25  & &\\
(4) & 0.70 & 1.09/0.82 & 1.08$^{4}$/0.71 & 0.51/1.08$^{4}$ & 0.13/0.14 & 0.15/0.15  & 0.36/0.34 & 1.34/2.20 & 0.57/0.25 & 0.75/0.42 & 0.66/0.29\\
 & & & & & & & &  2.81  & &\\
(4) & 0.26 & 0.92/0.85 & 0.95/0.92$^{4}$ & 0.92$^{4}$/1.49 & $^{3}$ & 0.18/0.08 & 0.31/0.34 & 1.89/1.60 & 0.55/0.41 & 0.70/0.44 & 0.63/0.40\\
 & & & & & & & &  2.79  & &\\
(4) & 0.57 & 3.03$^{1}$ & 2.07$^{1}$ & 0.24$^{1}$ & 0.69$^{1}$ & 0.50$^{1}$ & 0.67$^{1}$  & 3.45$^{1}$ & 0.91$^{1}$ & 1.19$^{1}$ & 1.08$^{1}$\\
 & & & & & & & &  2.70  & &\\
(4) & 0.62 & 1.40/0.71 & 0.94$^{4}$/0.97 & 0.57/0.94$^{4}$ & $^{3}$ & $^{3}$/0.19 & 0.36/0.28 & 1.19/2.26 & 0.57/0.28 & 0.60/0.41 & 0.56/0.43\\
 & & & & & & & & 2.58  & &\\
(4) & 0.66 &  $^{3}$ & $^{3}$  & $^{3}$ & $^{3}$ & 0.15/0.21 & 0.30/0.27 & 1.19/2.06 & 0.60/0.31 & 0.72/0.46 & 0.58/0.36\\
 & & & & & & & &  2.59  & &\\
(4) & 0.87 & 3.12$^{1}$ & 2.14$^{1}$ & 0.21$^{1}$ & 0.31$^{1}$ & 0.37$^{1}$ & 0.71$^{1}$ &  3.47$^{1}$ & 0.85$^{1}$ & 1.25$^{1}$ & 1.05$^{1}$\\
 & & & & & & & &  2.78  & &\\
(4) & 0.91 & 3.04$^{1}$ & 1.90$^{1}$ & 0.33$^{1}$ & 0.35$^{1}$ & 0.24$^{1}$ & 0.60$^{1}$ & 3.31$^{1}$ & 0.86$^{1}$ & 1.16$^{1}$ & 0.95$^{1}$\\
 & & & & & & & &  2.97  & &\\
(4) & 0.95 & $^{3}$  & 1.34$^{1}$ & 0.31$^{1}$ & $^{3}$ & 0.20$^{1}$ & 0.63$^{1}$ & 3.46$^{1}$ & 0.90$^{1}$ & 1.15$^{1}$ & 0.95$^{1}$\\
 & & & & & & & &  3.03  & &\\
(4) & 0.18 & 0.91/0.76 & 1.09/0.98$^{4}$ & 0.98$^{4}$/1.12 & $^{3}$ & 0.16/0.16 & 0.22/0.31 & 1.86/1.67 & 0.58/0.40 & 0.70/0.49 & 0.68/0.38 \\
 & & & & & & & &  3.07  & &\\
(4) & 0.22 & 1.52/1.37 & 1.33/0.99$^{4}$ & 0.99$^{4}$/1.70 & 0.16/0.24 & 0.17/0.29  & 0.27/0.31 & 2.09/1.59 & 0.59/0.41 & 0.80/0.53 & 0.69/0.43\\
 & & & & & & & & 3.19  & &\\
(4) & 0.27 & 1.25/1.24 & 1.08/0.96$^{4}$ & 0.96$^{4}$/0.36 & 0.17/0.15 & 0.18/0.23 & 0.31/0.30 & 1.99/1.74 & 0.57/0.42 & 0.75/0.53 & 0.64/0.43\\
 & & & & & & & &  3.10  & &\\
(4) & 0.48 & 3.30$^{1}$ & 1.97$^{1}$ & 0.44$^{1}$ & 0.50$^{1}$ & 0.29$^{1}$ & 0.67$^{1}$ & 3.41$^{1}$ & 0.94$^{1}$ & 1.28$^{1}$ & 1.12$^{1}$\\
 & & & & & & & &  2.91  & &\\
(4) & 0.56 & 2.67$^{1}$ & 2.22$^{1}$ & 0.28$^{1}$ & 0.36$^{1}$ & 0.44$^{1}$ & 0.70$^{1}$  & 3.55$^{1}$ & 0.99$^{1}$ & 1.36$^{1}$ & 1.18$^{1}$\\
 & & & & & & & &  2.91  & &\\
(5) & 0.29 & 1.33/1.20 & 0.77/1.66$^{4}$ & 1.66$^{4}$ & $^{3}$ & $^{3}$ & 0.26/0.26 &  2.04/1.56 & - & 0.91/0.43 & 0.34/0.46 \\
 & & & & & & & &  3.16  & &\\
(5) & 0.27 & 1.16/1.02 & 0.54/0.66$^{4}$ & 0.66$^{4}$ & $^{3}$ & $^{3}$ & 0.36/0.28  & 1.88/0.59 & - & 0.60/0.64 & 0.51/0.40 \\
 & & & & & & & &  2.87  & &\\
(5) & 0.64 & 1.09/0.59 & 0.89$^{4}$/0.81 & $^{4}$/0.35 & $^{3}$ & $^{3}$ & 0.30/0.24 & 2.05 & - & 0.65/0.26 & 0.52/0.22 \\
 & & & & & & & &  2.87  & &\\
(5) & 0.21 & 0.18/1.00 & 1.10/0.96$^{4}$ & $^{4}$/0.48 & 0.51$^{1}$ & 0.14/0.11 & 0.41/0.21 & 1.89/1.83 & - & 0.59/0.42 & 0.58/0.40 \\
 & & & & & & & &  3.19  & &\\
\noalign{\smallskip}
\hline
\end{tabular}
\end{flushleft}
{\scriptsize
H$\alpha$$_{i}$: The integrated total H$\alpha$ $EW$s value of both
components.;
$^{1}$ Data for primary and secondary components not deblended.;
$^{2}$ Mean value of two apertures in each spectrum or higher $S/N$
aperture measure;
$^{3}$ Data not measured due to very low $S/N$;
$^{4}$ These are the blended value of H line of one component with H$\epsilon$
 line from the other component;
$^{5}$ Observing run identification (see Sect. 2).
}
\end{table*}
}
\onltab{8}{
\begin{table*}
\caption[]{Emission fluxes
\label{tab:flux}}
\begin{center}
\scriptsize
\begin{tabular}{lcccccccccc}
\noalign{\smallskip}
\hline \hline
\noalign{\smallskip}
 &  \multicolumn{10}{c}{log$F$$_{\rm S}$}  \\
\cline{2-11}
\noalign{\smallskip}
 Obs. &
 \multicolumn{2}{c}{Ca {\sc ii}} & & & & & &
\multicolumn{3}{c}{Ca {\sc ii} IRT}  \\
\cline{2-3}\cline{9-11} \noalign{\smallskip} Idt.$^{5}$ &
 K   & H  & H$\epsilon$ & H$\delta$ & H$\gamma$ & H$\beta$ & H$\alpha$ &
$\lambda$8498 & $\lambda$8542 & $\lambda$8662
\scriptsize
\\
\noalign{\smallskip}
\hline
\noalign{\smallskip}
(1) & - & - & - & - & - & - & 6.89/7.28 & 6.41/6.13 & 6.59/6.38 & 6.55/6.22\\
(1) & - & - & - & - & - & - & 7.50$^{1}$ & 6.67$^{1}$ & 6.88$^{1}$ & 6.76$^{1}$\\
(1) & - & - & - & - & - & - & 7.13/7.36 & 6.63/6.54 & 6.94/6.84 & 6.61/6.51\\
(1) & - & - & - & - & - & - & 7.04/7.20 & - & - & -\\
(1) & - & - & - & - & - & - & 7.49$^{1}$ & 6.60$^{1}$ & 6.81$^{1}$ & 6.69$^{1}$\\
(1) & - & - & - & - & - & - & 7.10/7.24 & 6.45/6.11 & 6.65/6.41 & 6.55/6.20\\
(1) & - & - & - & - & - & - & 6.90/7.25 & 6.50/6.77 & 6.77/6.67 & 6.59/6.22\\
(1) & - & - & - & - & - & - & 6.94/7.22 & $^{3}$ & $^{3}$ & $^{3}$\\
(1) & - & - & - & - & - & - & 7.04/7.23 & 6.44/6.13 & 6.72/6.52 & 6.49/6.26\\
(1) & - & - & - & - & - & - & 7.04/7.58 & 6.63$^{1}$ & 6.84$^{1}$ & 6.59$^{1}$\\
(2) & - & - & - & - & - & 6.40/6.20 & 7.23/7.33 & - & 6.72/6.28 & 6.63/6.28\\
(2) & - & - & - & - & - & 6.68$^{1}$ & 7.48$^{1}$ & - & 6.90$^{1}$ & 6.84$^{1}$\\
(2) & - & - & - & - & - & 6.85$^{1}$ & 7.15/7.38 & - & 6.74/6.38 & 6.73/5.92\\
(2) & - & - & - & - & - & 6.72$^{1}$ & 7.49$^{1}$ & - & 6.93$^{1}$ & 6.88$^{1}$
\\
(2) & - & - & - & - & - & 6.74$^{1}$ & 7.54$^{1}$ & - & 6.92$^{1}$ & 6.85$^{1}$\\
(2) & - & - & - & - & - & 6.96$^{1}$ & 7.66$^{1}$ & - & 6.97$^{1}$ & 6.93$^{1}$\\
(2) & - & - & - & - & - & 6.73$^{1}$ & 7.35/7.09 & - & 6.70/6.25 & 6.64/6.41\\
(2) & - & - & - & - & - & 6.86$^{1}$ & 7.48/7.09 & - & 6.97/6.87 & 7.00/6.90\\
(3) & 7.15/7.09 & 7.06/7.13 & $^{3}$ & $^{3}$ & 5.73/5.94 & 6.57/6.38 & 7.28/7.19 & 6.77/6.55 & 6.61/6
.40 & 6.70/6.40\\
(3) & 7.43$^{1}$ &  7.28$^{1}$  & 6.69$^{1}$ & 6.17$^{1}$ & 6.30$^{1}$ & 6.77$^{1}$ & 7.54$^{1}$ & 6.9
5$^{1}$ & 6.84$^{1}$ & 6.89$^{1}$\\
(3) & 7.18/6.90 & 7.01/6.79 & 6.67/$^{3}$ & 5.83/5.82 & 5.96/6.00 & 6.56/6.16 &
7.07/7.19 & 6.83/6.35 & 6.67/6.11 & 6.73/6.28\\
(3) & 7.17/7.06 & 7.08/6.84 & 6.51/5.64 & 5.87/5.86 & 5.73/6.03 & 6.23/6.49 & 7.21/7.15 & 6.76/6.58 &
6.56/6.36 & 6.73/6.41\\
(4) & 6.93/6.71 & 6.93/6.65 & 6.60/$^{3}$ & 5.99/5.96 & 6.06/6.00 & 6.46/6.37 & 7.07/7.23 & 6.63/6.17
& 6.75/6.40 & 6.69/6.23\\
(4) & 6.86/6.73 & 6.87/6.76 & $^{3}$/6.97 & $^{3}$ & 6.14/5.73 & 6.40/6.37 & 7.22/7.10 & 6.61/6.38 & 6
.72/6.42 & 6.67/6.37\\
(4) & 7.38$^{1}$ & 7.21$^{1}$ & 6.28$^{1}$ & 6.71$^{1}$ & 6.58$^{1}$ & 6.73$^{1}$ & 7.48$^{1}$ & 6.83$
^{1}$ & 6.95$^{1}$ & 6.90$^{1}$\\
(4) & 7.04/6.65 &6.872/6.79 & 6.65/$^{3}$ & $^{3}$ & $^{3}$/6.10 & 6.46/6.28 &
7.02/7.25 & 6.63/6.22 & 6.65/6.38 & 6.62/6.41\\
(4) & $^{3}$ & $^{3}$ & $^{3}$ & $^{3}$ & 6.06/6.15 & 6.38/6.27 & 7.02/7.20 & 6.65/6.26 & 6.73/6.43 & 6.63/6.33\\
(4) & 7.39$^{1}$ & 7.23$^{1}$ & 6.22$^{1}$ & 6.37$^{1}$ & 6.45$^{1}$ & 6.76$^{1}$ & 7.48$^{1}$ & 6.80$^{1}$ & 6.97$^{1}$ & 6.89$^{1}$\\
(4) & 7.38$^{1}$ & 7.17$^{1}$ & 6.41$^{1}$ & 6.42$^{1}$ & 6.26$^{1}$ & 6.68$^{1}$ & 7.46$^{1}$ & 6.80$^{1}$ & 6.93$^{1}$ & 6.85$^{1}$\\
(4) & $^{3}$ & 7.02$^{1}$ & 6.39$^{1}$ & $^{3}$ & 6.18$^{1}$ & 6.70$^{1}$ & 7.48$^{1}$ & 6.82$^{1}$ & 6.93$^{1}$ &  6.85$^{1}$\\
(4) & 6.85/6.68 & 6.93/6.79$^{4}$ & $^{3}$/6.85 & $^{3}$ & 6.09/6.03 & 6.25/6.33 & 7.21/7.11 & 6.63/6.37 & 6.72/6.46 & 6.70/6.35\\
(4) & 7.08/6.94 & 7.02/6.79$^{4}$ & $^{3}$/7.03 & 6.08/6.20 & 6.11/6.29 & 6.34/6.33 & 7.26/7.09 & 6.64/6.38 & 6.77/6.50 & 6.71/6.41\\
(4) & 6.99/6.89 & 6.93/6.78$^{4}$ &$^{3}$/6.35 & 6.10/5.99 & 6.14/6.18 & 6.40/6.31 & 7.24/7.13 & 6.62/6.40 & 6.75/6.50 & 6.68/6.41\\
(4) & 7.41$^{1}$ & 7.19$^{1}$ & 6.54$^{1}$ & 6.57$^{1}$ & 6.35$^{1}$ & 6.73$^{1}$ & 7.48$^{1}$ & 6.84$^{1}$ & 6.98$^{1}$ & 6.92$^{1}$\\
(4) & 7.32$^{1}$ & 7.24$^{1}$ & 6.34$^{1}$ & 6.43$^{1}$ & 6.53$^{1}$ & 6.75$^{1}$ & 7.49$^{1}$ & 6.87$^{1}$ & 7.00$^{1}$ & 6.94$^{1}$\\
(5) & 6.78/7.02 & 7.02/6.88$^{4}$ & 7.12/7.02 & $^{3}$/$^{3}$ & $^{3}$/$^{3}$ & 6.32/6.25 & 7.25/7.08 & - & 6.83/6.41 & 6.40/6.43\\
(5) & 6.63/6.62 & 6.96/6.81$^{4}$ & 6.71/$^{3}$ & $^{3}$/$^{3}$ & $^{3}$/$^{3}$ & 6.46/6.28 & 7.22/6.66 &  - & 6.65/6.58 & 6.58/6.37\\
(5) & 6.84/6.71 & 6.93$^{4}$/6.57 & $^{4}$/6.34 & $^{3}$/$^{3}$ & $^{3}$/$^{3}$ & 6.38/6.21 & 7.00/7.20 & - & 6.68/6.19 & 6.59/6.11\\
(5) & 6.94/6.78 & 6.15/6.80$^{4}$ & 6.90/6.48 & 6.58$^{1}$ & 6.03/5.86 & 6.52/6.16 & 7.22/7.15 &  - & 6.49/6.40 & 6.63/6.37\\
\noalign{\smallskip}
\hline
\end{tabular}
\end{center}
{\scriptsize Notes as in previous Table.}
\end{table*}
}


\begin{thebibliography}{}
%
\bibitem[1992]{} Applegate, J. H. 1992, ApJ, 365, 621

\bibitem[1985]{} Barden, S. C.
1985, ApJ, 295, 162

\bibitem[1979]{} Beavers, W. I., Eitter, J. J., Ketelsen, D. A.,
 \& Oesper, D. A.
1979, PASP, 91, 698

\bibitem[1989]{} Buzasi, D. L. 1989, PhD Thesis, Pennsylvania State Univ.


\bibitem[1991]{} Chester, M. M.
1991, PhD Thesis, Pennsylvania State Univ.

\bibitem[2004]{} Cumming, A. 2004, MNRAS, 354, 1165

\bibitem[1984a]{} Eggen, O. J.
1984, ApJS, 55, 597

\bibitem[1989]{} Eggen, O. J.
1989, PASP, 101, 366

\bibitem[1997]{} ESA 1997, The Hipparcos and Tycho Catalogues, ESA SP-1200

\bibitem{} Favata, F., Barbera, M., Micela, G., \& Sciortino, S. 1993, A\&A, 277, 428


\bibitem[1997]{} Fekel, F. C. 1997, PASP, 109, 514

\bibitem[2005]{} Frasca, A., $\&$ Lanza, A. F. 2005, ApJ, 429, 309

\bibitem[2005]{} G\'alvez, M. C. 2005, PhD Thesis, Universidad Complutense
 de Madrid

\bibitem[2002]{} G\'alvez, M. C., Montes, D., Fern\'{a}ndez-Figueroa M. J., L\'opez-Santiago, J., De Castro, E., \& Cornide, M. 2002, A\&AS, 389, 524
 (Paper IV)

\bibitem[]{} G\'alvez, M. C., Montes, D., Fern\'andez-Figueroa, M. J.,
De Castro, E., \& Cornide, M. 2006, SEA/JENAM 2004, The many scales in the
Universe, Joint European and National Astronomy Meeting, Springer, ISBN-10
1-4020-4351-1, J.C. Del Toro Iniesta, et al. (eds.), Session 3, CD P29


\bibitem[]{} G\'alvez, M. C., Montes, D., Fern\'andez-Figueroa, M. J.,
De Castro, E., \& Cornide, M. 2007, Proceedings of Binary Stars as
Critical Tools and Tests in Contemporary Astrophysics, IAU Symp.
240, 26th meeting of the IAU, Special Session 3, S240, \#214


\bibitem[2003]{} Garc\'{\i}a-\'Alvarez D, Foing B. H., Montes D., et al. 2003, A\&A, 397, 285

\bibitem[1992]{} Hall, J. C., \& Ramsey, L. W.
 1992, AJ, 104, 1942

\bibitem[1996]{} Hall, J. C.
1996, PASP, 108, 313

\bibitem[1995]{} Henry, G. W., Fekel, F. C., \& Hall D. 1995, AJ, 110, 2926

\bibitem[2000]{} H$\o$g, E., et al. 2000, A\&A, 355, L27

\bibitem[1995]{} Jeffries, R. D., Bertram, D., \& Spurgeon, B. R.
1995, MNRAS, 276, 397

\bibitem[1995]{} Kalimeris, A., Mitrou, C. K., Doyle, J. G., Antonopoulou, E., \& Rovithis-Livaniou, H. 1995, A\&A, 293, 371

\bibitem[1998]{} Lanza, A. F., Rodono, M., $\&$ Rosner, R. 1998, MNRAS, 296, 893
\bibitem[1999]{} Lanza, A. F., \& Rodon\'o, M. 1999, A\&A, 349, 887

\bibitem[2005]{} Lanza, A. F. 2005, MNRAS, 364, 238 L

\bibitem[2006]{} Lanza, A. F. 2006, MNRAS, 369, 1773

\bibitem[]{} L\'opez-Santiago, J., Montes D., Fern\'andez-Figueroa
M. J., \& Ramsey L. W. 2003, A\&A, 411, 489

\bibitem[1983]{} Matese, J. J., $\&$ Whirtmere, D. P. 1983, A$\&$A, 117, L7

\bibitem[1995]{} Mason, K. O., Hassall, B. J. M., Bromage, G. E., et al.
1995, MNRAS, 274, 1194

\bibitem[1995]{} Montes, D., Fern\'{a}ndez-Figueroa, M. J., De Castro, E.,
 \& Cornide, M.
1995, A\&A, 294, 165

\bibitem[1997]{} Montes, D., Fern\'{a}ndez-Figueroa, M. J., De Castro, E., \&
Sanz-Forcada, J. 1997, A\&AS, 125, 263 (Paper I)

\bibitem[1998a]{} Montes, D., Sanz-Forcada, J., Fern\'{a}ndez-Figueroa, M. J., De Castro, E., \& Poncet, A.  1998, A\&A,  330, 155 (Paper II)

\bibitem[2000]{} Montes, D., Fern\'andez-Figueroa, M. J.,
De Castro, E., Cornide, M., Latorre, A., \& Sanz-Forcada J.
2000, A\&AS, 146, 103 (Paper III)

\bibitem[2001a]{} Montes, D., L\'opez-Santiago, J., G\'alvez, M. C.,
Fern\'andez-Figueroa, M. J., De Castro, E., \& Cornide, M.
2001a, MNRAS, 328, 45

\bibitem[2001b]{} Montes, D., L\'opez-Santiago, J.,
Fern\'andez-Figueroa, M. J., \& G\'alvez, M. C. 2001b, A\&A, 379, 976

\bibitem{} Pfeiffer, M. J.,  Frank, C., Baumueller, D., et al.
1998, A\&AS, 130, 381

\bibitem[1986]{} Press, W. H., Flannery, B. P., \& Teukolsky, S. A.
1986, Numerical recipes. The art of scientific computing,
Cambridge: University Press

\bibitem{} Schmidt-Kaler, T. 1982, in Landolt-B\"{o}rnstein, Vol. 2b,
 ed K. Schaifers, \& H. H. Voig (Heidelberg: Springer)

\bibitem{} Soderblom, D. R., Oey, M.S., Johnson, D. R. H., \& Stone, R. P. S. 1990, AJ, 99, 595

\bibitem[2001]{} Sowell, J. R., Hughes, S. B., Hall, D.S., \& Howard, B. A.
2001, AJ, 122, 1965

\bibitem[2000]{} Strassmeier, K. G., Washuettl, A., Granzer, Th., Scheck, M., \& Weber, M. 2000, A\&AS, 142, 275

\bibitem[1979]{} Tonry, J., \&  Davis, M.
1979, AJ, 84 1511

\bibitem[1981]{} Tonry, J., \&  Davis, M. 1981, AJ, 246, 666

%
\end{thebibliography}
\end{document}